\documentclass[aps,pra,onecolumn,showpacs,superscriptaddress]{revtex4}
\usepackage{graphicx}
\usepackage{amsmath}
\usepackage{amssymb}

\input{epsf}
\begin{document}

\title{Aligned dipolar Bose-Einstein condensate in a double-well potential:
From cigar-shaped to pancake-shaped }

\author{M. Asad-uz-Zaman}
\affiliation{Department of Physics and Astronomy,
Washington State University,
  Pullman, Washington 99164-2814, USA}
\author{D. Blume}
\affiliation{Department of Physics and Astronomy,
Washington State University,
  Pullman, Washington 99164-2814, USA}

\date{\today}

\begin{abstract}
We consider a Bose-Einstein condensate (BEC),
which is characterized by
long-range and anisotropic dipole-dipole interactions and
vanishing $s$-wave scattering length,
 in a double-well
potential.
The properties of this system are investigated as functions
of the height of the barrier that
splits the harmonic trap into two halves, the
number of particles (or dipole-dipole strength)
and the aspect ratio $\lambda$,
which is defined as the ratio between the axial and longitudinal
trapping frequencies
$\omega_z$ and
$\omega_{\rho}$.
The  
phase diagram 
is determined by analyzing the stationary
mean-field solutions.
Three distinct regions are found:
a region where the energetically lowest lying stationary solution 
is symmetric, a 
region where the energetically lowest lying stationary solution 
is located asymmetrically in
one of the wells,
and a region where the system is mechanically unstable.
For sufficiently large aspect ratio $\lambda$ and sufficiently
high barrier height,
the system consists of two connected quasi-two-dimensional sheets
with density profiles whose maxima are located either at $\rho=0$ or away
from $\rho=0$.
The stability of the stationary solutions is investigated
by analyzing the Bogoliubov de Gennes excitation spectrum
and the dynamical response to small perturbations.
These studies reveal 
unique oscillation frequencies and distinct collapse mechanisms.
The results derived within the 
mean-field  framework are complemented by an
analysis based on a
two-mode model.
\end{abstract}


\maketitle
\section{Introduction}
\label{introduction}
Dipolar BECs have recently attracted a lot of
attention both theoretically and experimentally~\cite{bara08,laha08a}.
The experimental realization of a Cr BEC just a few years ago
constitutes an important milestone~\cite{grie05}. Compared to alkali atoms, 
Cr has a comparatively large magnetic dipole moment of
$6 \mu_B$, which leads to an enhancement of the dipole-dipole
interactions by a factor of 36 compared to alkali atoms.
The anisotropy of the dipole-dipole interactions has been
observed experimentally by analyzing time of flight expansion images
of Cr BECs released from a cylindrically symmetric external confining
potential~\cite{stuh05}. If combined with theoretical calculations,
the time of flight images reveal the initial
density distribution of the dipolar gas and depend,
e.g., on whether the magnetic
dipole moments are aligned along the axial or longitudinal 
confining directions, respectively.
Furthermore, by taking advantage
of the tunability of the $s$-wave scattering length near a 
magnetic Fano-Feshbach resonance, the relative 
importance of the dipole-dipole
interactions can be changed~\cite{wern05,koch08,laha08},
paving the way for a variety of interesting experimental studies.
Dipolar BECs are
characterized by intriguing collapse 
mechanisms~\cite{sant00,yi00,yi01,gora02,eber05,bort06,fisc06,rone07,dutt07,laha08,koch08,metz09}, and
unique excitation spectra~\cite{giov02,sant03,rone07} and vortex 
structures~\cite{coop05,zhan05,yi06,odel07,wils09}. 
In addition, 
dipolar gases loaded 
into optical lattices may allow for the realization of novel
phases~\cite{gora02a,dams03,meno07}.

Although the dipole-dipole interactions in alkali gases are 
too weak to result in observable effects in most experiments,
it is believed that they play a 
decisive role in the formation of spin textures in $^{87}$Rb 
condensates~\cite{sadl06,veng08},
in the dynamics of Bloch oscillations of $^{39}$K BECs loaded into
an optical lattice~\cite{fatt08} and 
in $^7$Li BECs~\cite{poll09}.
Furthermore, BECs and degenerate 
Fermi gases that consist of polar molecules
may be realized in the near future~\cite{ni08,deig08,danz08}. 
This prospect adds a new intriguing twist
since the interactions 
between two polar molecules can be tuned by an external
electric field~\cite{mari98}. This opens 
the possibility to enter the strongly-correlated 
regime and thus to realize a variety of condensed matter 
analogs~\cite{gora02a,lewe07,bloc08}.

This paper considers an aligned dipolar BEC in a double well geometry.
Double well potentials play an 
important role in chemical and condensed
matter physics, among other areas.
In the context of cold atom physics,
$s$-wave dominated alkali systems in a double-well 
have, e.g., been used to study
Josephson-type 
oscillations~\cite{aubr96,smer97,ande98a,ragh99,giov00a,cata01,albi05,anan06,gati07}. 
The density oscillations of the 
Bose gas can be interpreted as corresponding to the charge current
that characterizes ``standard'' condensed matter Josephson junctions.
Related to this, the macroscopic quantum 
self-trapping of atoms in one of the wells
has been demonstrated experimentally and has been interpreted within a 
two-mode model that can be derived from the Gross-Pitaevskii (GP) 
equation~\cite{albi05,anan06}.
The double well system has also been used to experimentally 
study spin-squeezing~\cite{este08}. 
In this context, the left and the right
wells of the system serve as
the two arms of an
interferometer~\cite{shin04}. The number difference and 
relative phase of the double well
system are conjugate
variables, whose combined measurements has revealed that the system is 
entangled~\cite{este08}.

Here, we investigate the behaviors of 
aligned dipoles under cylindrically symmetric 
harmonic confinement with a repulsive Gaussian potential
centered at $z=0$.
We limit ourselves
to situations where the dipoles are aligned along one
of the symmetry axis of the external harmonic confining potential. 
This restriction reduces the parameter space and also
significantly reduces the numerical efforts.
Arguably, it may be the conceptually
simplest case. 
We are particularly interested in determining the phase
or stability diagram for
both cigar-shaped and pancake-shaped harmonic confinement.
The boundaries of these phase diagrams are governed by
the non-trivial interplay of the dipole-dipole interactions, 
the energy due to the external
harmonic confinement, the energy due to the Gaussian potential
and the kinetic energy. 
The interplay of these energy contributions
leads to 
density profiles unique to anisotropic
interactions.
In agreement with Ref.~\cite{xion09}, we observe Josephson oscillations
as well as 
macroscopic quantum
self-trapping of the system for appropriately chosen
parameters. We characterize the transition between these two regimes
by analyzing the excitation spectrum and the real time response of the system
to a small perturbation.
Our study of two neighboring pancake-shaped dipolar gases can
be viewed as a first step towards understanding a
multi-layer system of dipolar
pancakes. 
For a single pancake, an angular roton instability has been predicted to
occur~\cite{rone07}. For 
a layer of two-dimensional dipolar BECs,
a new length scale is given by the interlayer distance and
the roton instability has been predicted to be enhanced compared
to the single layer case~\cite{wang08}. 
Other multi-layer studies can be found in Refs.~\cite{klaw09,kobe09}.

The remainder of this paper is organized as follows. 
Section~\ref{theory}
introduces the mean-field GP equation and
discusses how we determine the stationary and time-dependent
solutions. The Bogoliubov de Gennes equations that are employed to
determine the
excitation spectrum of the dipolar gas are introduced.
Section~\ref{twomode} reviews the two-mode model which provides
an intuitive understanding of the time-independent
and time-dependent GP solutions in the small $\lambda$ regime.
Section~\ref{stationary}
presents our stationary solutions. 
We discuss the
phase diagram as functions of the number of particles
(or equivalently, the mean-field strength), the aspect ratio
and, in selected cases, the barrier height. 
Section~\ref{timedependent} presents our time-dependent studies.
We investigate certain dynamically stable
regimes and deduce distinct
collapse mechanisms from the response of the system to a small
perturbation. In addition, selected Bogoliubov de Gennes eigenmodes
are discussed.
Lastly, Sec.~\ref{conclusion} summarizes
our main findings and discusses possible future studies.

\section{Mean-field description of dipolar BECs}
\label{theory}
Section~\ref{sec_gpground} introduces the time-dependent mean-field
GP equation for a dipolar BEC and discusses the numerical 
techniques employed to determine stationary
and time-dependent solutions. Section~\ref{sec_gpexcited} 
reviews the Bogoliubov de Gennes
equations for the dipolar BEC.

\subsection{Gross-Pitaevskii equation}
\label{sec_gpground}
The time-dependent GP equation for a dipolar 
BEC consisting of $N$ identical point dipoles is 
given by~\cite{yi00,gora00,yi01}
\begin{eqnarray}
\label{eq_gp}
i \hbar \frac{\partial \psi(\vec{r},t)}{\partial t} =
H
\psi(\vec{r},t),
\end{eqnarray}
where the mean-field Hamiltonian $H$
reads
\begin{eqnarray}
\label{eq_ham}
H=
- \frac{\hbar^2}{2m}\nabla^2 + V_{\mathrm{ext}}(\vec{r}) + \nonumber \\
(N-1) \int V_{\mathrm{dd}}(\vec{r}-\vec{r}') |\psi(\vec{r}',t)|^2 d^3 \vec{r}' .
\end{eqnarray}
Here, $m$ denotes the mass of the dipoles.
We interpret $\psi(\vec{r},t)$ as a single-particle wave function
and correspondingly use the normalization
$\int |\psi(\vec{r},t)|^2 d^3\vec{r} = 1$.
The external cylindrically-symmetric confining potential $V_{\mathrm{ext}}$
consists of a harmonic trapping potential $V_{\mathrm{ho}}$
with angular frequencies
$\omega_{\rho}$ and $\omega_z$
and a Gaussian barrier $V_{\mathrm{g}}$
with height $A$ ($A>0$) and width $b$,
\begin{eqnarray}
V_{\mathrm{ext}}(\vec{r}) = V_{\mathrm{ho}}(\rho,z) + V_{\mathrm{g}}(z),
\end{eqnarray}
where
\begin{eqnarray}
\label{eq_vho}
V_{\mathrm{ho}}(\rho,z) = \frac{1}{2} m (\omega_{\rho}^2 \rho^2 + \omega_z^2 z^2)
\end{eqnarray}
and
\begin{eqnarray}
\label{eq_vgauss}
V_{\mathrm{g}}(z) = A \exp \left(- \frac{z^2}{2 b^2} \right).
\end{eqnarray}
We define the aspect ratio $\lambda$ of the harmonic confining potential
as
\begin{eqnarray}
\label{eq_lambda}
\lambda = \frac{\omega_z}{\omega_{\rho}}.
\end{eqnarray}
Throughout, we employ cylindrical coordinates and write
$\vec{r}=(\rho, \varphi,z)$.

The third term on the right hand side of Eq.~(\ref{eq_ham})
represents the mean-field
potential, which depends on both the density of the system
and
the dipole-dipole potential $V_{\mathrm{dd}}$. Throughout, we assume that the
dipoles are aligned along the $z$-axis,
\begin{eqnarray}
\label{eq_vdd}
V_{\mathrm{dd}} (\vec{r}-\vec{r}') = 
d^2 \frac{1 - 3 \cos^2 \vartheta}{|\vec{r}-\vec{r}'|^3},
\end{eqnarray}
where $d$ denotes the strength of the
dipole moment of the dipolar atom or molecule
under study and $\vartheta$ the angle between the relative distance vector
$\vec{r}-\vec{r}'$ and the $z$-axis. Throughout, we assume that 
the
$s$-wave scattering length $a_s$ vanishes, implying the absense of
the usual $s$-wave contact interaction
term in Eq.~(\ref{eq_ham}).
For 
dipolar Cr BECs,
e.g., this can be achieved by varying an external magnetic field in
the vicinity of a Fano-Feshbach resonance~\cite{wern05,koch08,laha08}.
 
Rewriting the 
integro-differential equation [Eq.~(\ref{eq_gp}) with Eqs.~(\ref{eq_ham})-(\ref{eq_vdd})]
in harmonic oscillator units $a_z$ and $E_z$,
where
\begin{eqnarray}
\label{eq_aho}
a_z = \sqrt{\frac{\hbar}{m \omega_z}}
\end{eqnarray}
and
\begin{eqnarray}
\label{eq_ez}
E_z = \hbar \omega_z,
\end{eqnarray}
shows that the GP equation
depends on four dimensionless 
parameters: i) $d^2(N-1)/(E_z a_z^3)$, which 
characterizes the strength of the mean-field potential;
ii) the aspect ratio $\lambda$;
iii) the scaled barrier height $A/E_z$;
and
iv) the scaled barrier width $b/a_z$.
To reduce the parameter space, we consider a 
fixed barrier width $b$, i.e., $b=a_z/5$. 
While most of our calculations are performed for
$A=12E_z$, we consider smaller barrier heights in selected cases.
The aspect ratio
is varied from $\lambda=0.1$ (cigar-shaped external harmonic confinement)
to $\lambda = 12$ (pancake-shaped external harmonic confinement).
Lastly,
the dimensionless mean-field strength $D$,
\begin{eqnarray}
D= \frac{d^2(N-1)}{E_z a_z^3},
\end{eqnarray}
 is, for a given $A$ and $\lambda$,
varied from 0 to the value 
$D_{{\mathrm{cr}}}$ at which collapse occurs.

In practice, the mean-field strength $D$ can be adjusted
by loading the double-well potential with 
condensates of varying particle number $N$. 
More conveniently, one might envision tuning the electric
dipole moment of a molecular BEC through
the application of an external electric field~\cite{mari98} or,
in the case of magnetic Cr BECs, by changing the ratio between the 
dipole-dipole
and the $s$-wave interactions through the application of an external magnetic
field in the vicinity of a Fano-Feshbach resonance~\cite{wern05,koch08,laha08}.
Although our study considers $a_s=0$ and varying $D$,
the latter scenario should allow for the observation of a number
of features
predicted in this study.
Experimentally, the Gaussian barrier potential of varying height and width
can be realized by a repulsive dipole beam with adjustable
intensity and waist.

The solutions to the 
integro-differential mean-field equations 
have to be determined self-consistently
since the density $|\psi|^2$, which is part of the solution sought, 
also enters into the mean-field potential. The stationary
solutions can be written as 
$\psi(\vec{r})= \Psi(\rho,z) h(\varphi)$ 
with $h(\varphi)=\exp(i k \varphi)/\sqrt{2 \pi}$.
In the following,
we seek stationary 
solutions with azimuthal 
quantum number $k=0$.
Our calculation of the excitation spectrum do, however, 
include $k>0$ modes (see Sec.~\ref{sec_gpexcited}). 
The evaluation of the integral contained in the
mean-field potential can be performed most readily
by transforming to momentum space via a combined
Fourier-Hankel transform~\cite{rone06a}.
To determine the stationary solutions, we implemented two
different approaches: i) 
We minimize the total
energy of the system following the conjugate gradient 
approach~\cite{modu03}. 
In this approach, the solution is expanded
in terms of harmonic oscillator basis functions in $\rho$ and $z$,
and the expansion coefficients are optimized so as to minimize the 
total energy per particle.
ii) We propagate an initial state in imaginary time till
the stationary solution has been projected out.

The basis functions and the initial state are both
represented on a grid in the $\rho$ and
$z$ directions. The grid along $\rho$ is chosen according to the
zeroes of the Bessel functions (see Ref.~\cite{rone06a}),
which are
distributed roughly linearly. Along the $z$-direction, 
we use a linear grid.
For most calculations, a grid of $N_{\rho} \times N_z =
64 \times 128$ is sufficient.  
We employ a rectangular simulation box
of lengths $[0,\rho_{\mathrm{max}}]$ and $[-z_{\mathrm{max}},z_{\mathrm{max}}]$. 
For pancake-shaped systems (i.e., $\lambda>1$),
a ``cutoff'' is used for the dipolar potential (i.e., the interaction
is truncated for $|z|>z_{\mathrm{max}}$), which
reduces the interaction of the true BEC with an ``artificial image BEC'' and
thus allows for the usage of
a smaller $z_{\mathrm{max}}$~\cite{rone06a}. 
For $\lambda < 1$, no cutoff is employed.
Typical values for $\rho_{\mathrm{max}}$ and $z_{\mathrm{max}}$ are around
$15 a_{\rho}$ and $12a_z$, respectively,
where
$a_{\rho} = \sqrt{\hbar/(m \omega_{\rho})}$.

We have checked that the conjugate gradient 
and imaginary time evolution approaches result,
within our
numerical accuracy,
in identical energies and densities. 
Furthermore, for
vanishing barrier height, i.e., for $A=0$,
our solutions 
for 
cylindrically symmetric harmonic traps
agree with those reported in the literature~\cite{rone06a,rone07}. 
For non-vanishing barrier height, we compared our solutions with
those reported in Ref.~\cite{xion09}. 
Our energies and chemical potentials 
are in reasonable agreement with
those reported in Ref.~\cite{xion09}.
For $A=4E_z$, $b=0.2a_z$, $\lambda=0.1$ and $D=0.6$, e.g.,
we find $E/N=10.69E_z$ and $\mu=10.00E_z$
while the values reported in Fig.~2 of Ref.~\cite{xion09} 
are smaller by about $3$ and $4$\%, respectively. These deviations 
are somewhat larger than our estimated numerical uncertainty.

The time dynamics of the system
is determined by evolving a given initial state in real time.
The initial state is chosen according to the 
variational two-mode model wave function
(see Sec.~\ref{twomode}) or by adding a small random or smooth perturbation
to the stationary GP wave function
of the energetically lowest lying state.
If the system collapses to a high density state 
in response to the application of a small perturbation, 
then our simulations are only  able
to follow the real time evolution for a limited time period.
Eventually, our grid becomes too coarse to accurately
present the time evolved state. Since our main aim is directed
at identifying the stability and
the collapse mechanisms, this artefact does not pose any true
limitations on our analysis. In fact, once the density becomes 
sufficiently high, the mean-field GP description breaks down anyways and
beyond mean-field corrections need to be included.
Such a treatment is, however, beyond the scope of the present work.

\subsection{Bogoliubov de Gennes equations}
\label{sec_gpexcited}
In addition to time-evolving a given initial state, we 
analyze the stability of the dipolar BEC by seeking solutions to the
time-dependent GP equation of the form~\cite{dalf98}
\begin{eqnarray}
\label{eq_bdg}
\psi(\vec{r},t) = \exp(-i \mu t / \hbar) \left[ \psi_0(\vec{r})
+
\delta \psi(\vec{r},t) \right],
\end{eqnarray}
where $\psi_0(\vec{r})$ denotes the energetically lowest lying 
solution of
the time-independent GP equation 
with $k=0$ and $\mu$ the corresponding chemical potential.
We seek ``perturbations'' $\delta \psi(\vec{r},t)$
that oscillate with frequency $\omega$, 
\begin{eqnarray}
\label{eq_bdg2}
\delta \psi(\vec{r},t) = 
u(\vec{r}) \exp(-i \omega t) 
+ v^*(\vec{r}) \exp( i \omega t),
\end{eqnarray}
where $u(\vec{r})$ and $v(\vec{r})$ 
denote the Bogoliubov de Gennes ``particle'' and ``hole'' 
functions~\cite{dalf97}.
Plugging Eq.~(\ref{eq_bdg}) with $\delta \psi$ given by Eq.~(\ref{eq_bdg2})
into Eq.~(\ref{eq_gp}), keeping terms up to first order in 
$\delta \psi(\vec{r},t)$ and its complex conjugate,
and equating the coefficients of the terms oscillating with
$\exp(-i\omega t)$ and $\exp(i \omega t)$, respectively,
we find the Bogoliubov de Gennes
equations~\cite{rone06a}
\begin{eqnarray}
\label{eq_bdg3}
\hbar \omega u(\vec{r}) = 
{\cal{A}}(\vec{r}) u (\vec{r}) + 
\nonumber \\
(N-1) \int V_{\mathrm{dd}}(\vec{r}-\vec{r}') \psi_0^*(\vec{r}') u(\vec{r}')  
d^3\vec{r}' \psi_0(\vec{r}) + \nonumber \\
(N-1) \int V_{\mathrm{dd}}(\vec{r}-\vec{r}') 
\psi_0(\vec{r}') v(\vec{r}')   d^3 \vec{r}' \psi_0(\vec{r})
\end{eqnarray}
and
\begin{eqnarray}
\label{eq_bdg4}
-\hbar \omega v^*(\vec{r}) = 
{\cal{A}}(\vec{r}) v^*(\vec{r}) 
+ \nonumber \\
(N-1) \int V_{\mathrm{dd}}(\vec{r}-\vec{r}') \psi_0^*(\vec{r}') v^*(\vec{r}')  
d^3\vec{r}' \psi_0(\vec{r}) + \nonumber \\
(N-1) \int V_{\mathrm{dd}}(\vec{r}-\vec{r}') 
\psi_0(\vec{r}') u^*(\vec{r}')   d^3 \vec{r}' \psi_0(\vec{r}).
\end{eqnarray}
In Eqs.~(\ref{eq_bdg3}) and (\ref{eq_bdg4}), 
the operator ${\cal{A}}(\vec{r})$
is defined as
\begin{eqnarray}
\label{eq_bdg6}
{\cal{A}}(\vec{r}) = H_0 - \mu + \nonumber \\
(N-1) \int V_{\mathrm{dd}}(\vec{r}-\vec{r}') | \psi_0(\vec{r}')|^2 d^3\vec{r}',
\end{eqnarray}
where
$H_0$ denotes 
the Hamiltonian 
of the non-interacting system,
\begin{eqnarray}
H_0 = -\frac{\hbar^2}{2m} \nabla^2 + V_{\mathrm{ext}}(\vec{r}).
\end{eqnarray}

Equations
(\ref{eq_bdg3}) and (\ref{eq_bdg4}) can be decoupled by
introducing two new functions $f$ and $g$,
$f(\vec{r}) = u(\vec{r})+v(\vec{r})$ and
$g(\vec{r}) = -u(\vec{r})+v(\vec{r})$.
Assuming, without loss of generality, that $\psi_0(\vec{r})$
is real,
we find 
\begin{eqnarray}
\label{eq_bdg5a}
\hbar^2 \omega^2 f(\vec{r}) =
{\cal{A}}(\vec{r}) \left[ {\cal{A}}(\vec{r})
f(\vec{r}) \right] + \nonumber \\
2(N-1) 
{\cal{A}}(\vec{r}) \left[ 
\int 
f(\vec{r}') V_{\mathrm{dd}}(\vec{r}-\vec{r}') \psi_0(\vec{r}')  
d^3 \vec{r}' \psi_0(\vec{r})\right]
\end{eqnarray}
and
\begin{eqnarray}
\label{eq_bdg5b}
\hbar^2 \omega^2 g(\vec{r}) =
{\cal{A}}(\vec{r}) \left[ {\cal{A}}(\vec{r})
g(\vec{r})\right] + \nonumber \\
2(N-1) \int V_{\mathrm{dd}}(\vec{r}-\vec{r}') \psi_0(\vec{r}') 
{\cal{A}}(\vec{r}') 
g(\vec{r}') 
d^3 \vec{r}' \psi_0(\vec{r}) .
\end{eqnarray}
Following Ref.~\cite{rone06a},
we solve Eq.~(\ref{eq_bdg5b})
for the square of the 
Bogoliubov de Gennes excitation
frequency $\omega$ and the corresponding eigenvector $g(\vec{r})$
iteratively using the Arnoldi method.
Once $g(\vec{r})$ is determined, the eigenvector $f(\vec{r})$ can 
be obtained from the identity
\begin{eqnarray}
\label{eq_bdg7}
f(\vec{r})=-\frac{1}{\hbar \omega} {\cal{A}}(\vec{r}) g(\vec{r}).
\end{eqnarray}
The physical meaning of $f$ is elucidated by calculating
the density $|\psi(\vec{r},t)|^2$ up to first order in $\delta \psi$
and its complex conjugate.
For real $u$ and $v$, this gives
\begin{eqnarray}
\label{eq_bdgdensity}
|\psi(\vec{r},t)|^2 \approx |\psi_0(\vec{r})|^2 + 
2 \cos(\omega t) \psi_0(\vec{r}) f(\vec{r}),
\end{eqnarray}
which
shows that $f(\vec{r})$,
together with $\psi_0(\vec{r})$ and $\omega$, 
determines
the time-dependent density.
Due to the cylindrical symmetry of the system, the $\varphi$ dependence 
of $f(\vec{r})$ separates,
$f(\vec{r})= \bar{f}(\rho,z) h(\varphi)$.
Section~\ref{timedependent} discusses the behavior of 
$\bar{f}(\rho,z)$,
which we refer to as the Bogoliubov de Gennes eigenmode,
for different  $k$ and various $(D, \lambda)$ combinations.

The  outlined approach allows for the determination of a sequence of excitation
frequencies for a given azimuthal quantum number $k$ at a time. 
It can be seen from Eq.~(\ref{eq_bdg2}) 
that a negative $\omega^2$ and thus a purely imaginary
$\omega$ corresponds to a situation where
the stationary
ground state solution is dynamically unstable.

\section{Two-mode model}
\label{twomode}
Atomic BECs, coupled through non-vanishing potential barriers, 
have been used extensively to model
coupled condensed matter systems such as 
$^3$He-B reservoirs~\cite{webb74,smer97,ragh99,giov00a,anan06}. 
Although neutral, the 
study of weakly-coupled atomic BECs allows, e.g., 
for the realization of a variety
of typical dc and ac effects that characterize charged Cooper pair
superconducting junctions~\cite{giov00a,levy07}.
The connection between weakly-coupled 
atomic BECs and more traditional condensed matter systems
becomes most apparent if the former is approximated by a two-mode
model and mapped to a Josephson like Hamiltonian. 
Here, our primary motivation for employing the two-mode model
is to
develop an intuitive understanding of some of the phenomena 
observed in our time-independent and time-dependent
mean-field studies.

Let $\psi_{\mathrm{S}}(\vec{r})$ and $\psi_{\mathrm{A}}(\vec{r})$
denote 
the energetically lowest lying stationary GP solutions that are 
respectively symmetric and anti-symmetric with respect to $z=0$.
If
the symmetric function $\psi_{\mathrm{S}}(\vec{r})$ is the energetically lowest lying 
solution of the stationary GP equation, we calculate it
by employing the conjugate gradient method or by evolving in
imaginay time (see Sec.~\ref{theory}).
The anti-symmetric solution $\psi_{\mathrm{A}}(\vec{r})$ is obtained by
restricting the basis functions employed in the conjugate gradient method
to functions that are anti-symmetric with respect to $z=0$.
Without loss of generality, we assume in the following
that $\psi_{\mathrm{S}}$ and $\psi_{\mathrm{A}}$ are real.
In the two-mode model,
the solutions $\psi_{\mathrm{S}}$ and $\psi_{\mathrm{A}}$ are treated as  a basis that 
defines
the two ``modes''
$\Phi_{\mathrm{L}}(\vec{r})$
and $\Phi_{\mathrm{R}}(\vec{r})$,
\begin{eqnarray}
\label{eq_basis}
\Phi_{L,R}(\vec{r}) = \frac{
\psi_{\mathrm{S}}(\vec{r}) \pm \psi_{\mathrm{A}}(\vec{r})}{\sqrt{2}}.
\end{eqnarray}
By construction, $\Phi_{L}(\vec{r})$ and $\Phi_{R}(\vec{r})$
are normalized to one and orthogonal to each other.
The functions $\Phi_{\mathrm{L}}$ and $\Phi_{\mathrm{R}}$ are, for appropriately chosen
parameters,
located 
predominantly in the left well and 
in the right well,
respectively.

Within the two-mode model, the time-dependent
wave function is approximated by (see, e.g., Ref.~\cite{dalf98})
\begin{eqnarray}
\label{eq_psitwomode}
\psi(\vec{r},t) = c_{\mathrm{L}}(t) \Phi_{\mathrm{L}}(\vec{r}) + 
c_{\mathrm{R}}(t) \Phi_{\mathrm{R}}(\vec{r}) ,
\end{eqnarray}
where the complex-valued time-dependent expansion coefficients
$c_{\mathrm{L}}(t)$ and $c_{\mathrm{R}}(t)$ are related through the normalization condition
$|c_{\mathrm{L}}(t)|^2 + |c_{\mathrm{R}}(t)|^2=1$.
Defining $c_{L,R}(t) = |c_{L,R}(t)| \exp[i \theta_{L,R}(t)]$,
the time evolution within the two-mode model
is governed by two variables: the fractional
difference $Z(t)$ 
of the population 
located in the left and in the right well,
\begin{eqnarray}
Z(t) = |c_{\mathrm{L}}(t)|^2 - |c_{\mathrm{R}}(t)|^2,
\end{eqnarray}
and the phase difference or relative phase $\phi(t)$,
\begin{eqnarray}
\phi(t) = \theta_{\mathrm{R}}(t) - \theta_{\mathrm{L}}(t).
\end{eqnarray}
Plugging Eq.~(\ref{eq_psitwomode})
into Eq.~(\ref{eq_gp}), multiplying by $\Phi_{\mathrm{L}}(\vec{r})$ and 
$\Phi_{\mathrm{R}}(\vec{r})$, respectively, and
integrating out the spatial degrees of freedom,
we obtain two coupled equations that govern the time dynamics:
\begin{eqnarray}
\label{eq_cc1}
i \hbar \frac{d{c}_{\mathrm{L}}(t)}{dt} = \nonumber \\
\left[E_0+B+ (U-B) |c_{\mathrm{L}}(t)|^2 \right]  c_{\mathrm{L}}(t)
- T c_{\mathrm{R}}(t)
\end{eqnarray}
and
\begin{eqnarray}
\label{eq_cc2}
i \hbar \frac{d{c}_{\mathrm{R}}(t)}{dt} = \nonumber \\
\left[E_0+ B+(U-B) |c_{\mathrm{R}}(t)|^2 \right]  c_{\mathrm{R}}(t)
- T c_{\mathrm{L}}(t).
\end{eqnarray}
In deriving Eqs.~(\ref{eq_cc1}) and (\ref{eq_cc2}),
we neglegted terms of the form
\begin{eqnarray}
(N-1) \times \nonumber \\
\int \int \Phi_i(\vec{r})\Phi_j(\vec{r})
V_{\mathrm{dd}}(\vec{r}-\vec{r}')
\Phi_k(\vec{r}')\Phi_l(\vec{r}') d^3 \vec{r}' d^3 \vec{r}
\end{eqnarray}
with $i \ne j$ or $k \ne l$, where 
$i,j,k$ and $l$ can take the
values $L$ and $R$.
These terms are small as long as $\Phi_{\mathrm{L}}$ and $\Phi_{\mathrm{R}}$
are located predominantly in the left well and in the right well,
respectively.
In Eqs.~(\ref{eq_cc1}) and (\ref{eq_cc2}),
the onsite, offsite (or interaction tunneling) and 
tunneling matrix elements $U$, $B$ and $T$ are defined as
\begin{eqnarray}
\label{eq_onsite}
U = 
(N-1) \times \nonumber \\
\int \int (\Phi_{\mathrm{L}}(\vec{r}))^2  V_{\mathrm{dd}}(\vec{r}-\vec{r}') 
(\Phi_{\mathrm{L}}(\vec{r}'))^2 d^3 \vec{r}' d^3\vec{r},
\end{eqnarray}
\begin{eqnarray}
\label{eq_offsite}
B = (N-1) \times \nonumber \\
\int \int (\Phi_{\mathrm{L}}(\vec{r}))^2  V_{\mathrm{dd}}(\vec{r}-\vec{r}') 
(\Phi_{\mathrm{R}}(\vec{r}'))^2 d^3 \vec{r}' d^3\vec{r},
\end{eqnarray}
and 
\begin{eqnarray}
\label{eq_tunneling}
T = 
\int \left[ 
\frac{-\hbar^2}{2m} \nabla \Phi_{\mathrm{L}}(\vec{r}) \cdot
\nabla \Phi_{\mathrm{R}}(\vec{r}) 
- \right. \nonumber \\
\left. \Phi_{\mathrm{L}}(\vec{r}) V_{\mathrm{ext}}(\vec{r}) \Phi_{\mathrm{R}}(\vec{r})
\right] d^3 \vec{r},
\end{eqnarray}
and
the ``zero point energy'' $E_0$ is defined as
\begin{eqnarray}
\label{eq_onen}
E_0 = \int \left[ \frac{\hbar^2}{2m} 
| \nabla \Phi_{\mathrm{L}}(\vec{r})|^2 + V_{\mathrm{ext}}(\vec{r}) 
(\Phi_{\mathrm{L}}(\vec{r}))^2 \right] d^3 \vec{r}.
\end{eqnarray}
Usage of $\Phi_{\mathrm{R}}(\vec{r})$ instead of $\Phi_{\mathrm{L}}(\vec{r})$
in Eqs.~(\ref{eq_onsite}) and 
(\ref{eq_onen}) gives the same result.

Rewriting the coupled equations (\ref{eq_cc1}) and (\ref{eq_cc2})
in terms of 
$Z(t)$ and $\phi(t)$
leads to the classical Hamiltonian $H_{TM}$
(using $\hbar =1$),
\begin{eqnarray}
\label{eq_hamtwomode}
H_{TM} = 2T \left[ \Lambda \frac{Z^2(t)}{2} - \sqrt{1- Z^2(t)} \cos(\phi(t))
\right],
\end{eqnarray}
where
\begin{eqnarray}
\label{eq_biglambda}
\Lambda = \frac{U-B}{2T}.
\end{eqnarray}
Notably,
$Z(t)$ and $\phi(t)$ are 
conjugate variables of the classical Hamiltonian.
The energy of $H_{TM}$ is conserved and can, e.g., be obtained by
inserting $Z(0)$ and $\phi(0)$ into Eq.~(\ref{eq_hamtwomode}).
The properties of $H_{TM}$ have been discussed in detail in the 
literature~\cite{smer97,ragh99,giov00a,anan06}.
Here, we review a few points that will aid in the understanding of our 
GP solutions.

The two-mode model immediately leads to three different
classes of stationary solutions, i.e., solutions with constant
$Z(t)$ and $\phi(t)$:
A symmetric solution for $\phi(t) = 2 \pi n$ ($n$ integer)  
and $Z(t)=0$; its energy is $-2T$.
An anti-symmetric solution for $\phi(t)= (2n+1) \pi$ ($n$ integer)
and $Z(t)=0$; its energy is $2 T$.
A symmetry-broken solution 
for $\phi(t) = (2n+1)\pi$
($n$ integer) and $Z(t) = \pm \sqrt{1- \Lambda^{-2}}$;
this solution exists only if $|\Lambda|>1$
and its energy is
$T ( \Lambda + \Lambda^{-1})$.
Section~\ref{stationary}
compares these stationary two-mode 
model solutions with those obtained from the
stationary GP solutions.

It turns out that Hamilton's equations of motion can be 
solved analytically for $H_{TM}$~\cite{ragh99}.
Of particular interest for our study is the so-called
Josephson oscillation frequency $\omega_{\mathrm{J}}$, which---for small
amplitude motion---can be expressed in terms 
of $\Lambda$ (with $\hbar$ ``restored''),
\begin{eqnarray}
\label{eq_omegaj}
\hbar \omega_{\mathrm{J,TM}} = 2T \sqrt{1 + \Lambda} .
\end{eqnarray}
Section~\ref{timedependent}
compares the two-mode model 
frequency $\omega_{\mathrm{J,TM}}$ with the frequency obtained
from the real time dynamics
and by solving the Bogoliubov de Gennes equations.
For the real time dynamics, we prepare
an initial state at time $t=0$ 
according to Eq.~(\ref{eq_psitwomode}) 
and then time evolve this state according to the time dependent
mean-field Hamiltonian.
A Fourier analysis of the expectation value of $z(t)$ then reveals the 
predominant excitation frequency.

\section{Discussion of stationary solutions}
\label{stationary}
This section discusses our solutions to the stationary
GP equation. 
In particular, we present the phase diagram 
as a function of the aspect ratio $\lambda$ and the mean-field strength
$D$ for a fixed barrier
height $A$ and discuss selected density profiles.
Furthermore, we discuss how the phase diagram changes 
with varying barrier height
and explain some of the GP results within the two-mode
model.

\begin{figure}
\vspace*{.2cm}
\includegraphics[angle=0,width=75mm]{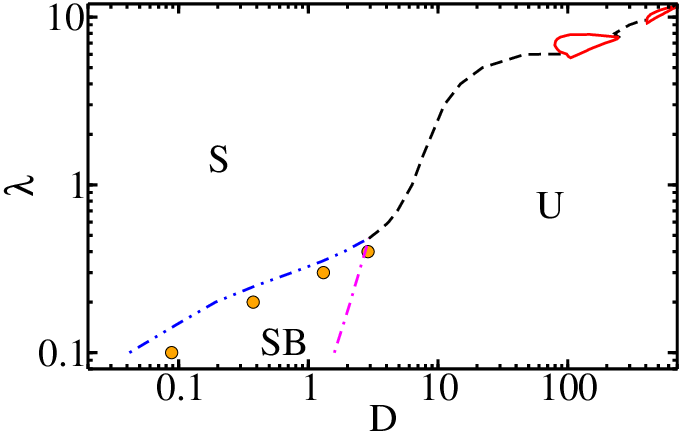}
\caption{
(Color online)
Character of the 
energetically lowest lying stationary GP solution 
for $b=0.2a_z$ and
$A=12E_z$:
The dash-dot-dotted, dash-dotted and dashed lines indicate those $\lambda$ and
$D$ values at which the character
of the energetically lowest lying stationary GP solution 
with $k=0$ changes from
symmetric (S) to symmetry-broken (SB),
from symmetry-broken to unstable (U),
and from symmetric to unstable, respectively.
The ``solid (red) islands'' in the upper right corner of the phase
diagram (comparatively large $D$ and $\lambda$) indicate 
two regions of the phase diagram where the solutions are symmetric
but where the density maximum is located at $\rho>0$;
these islands are discussed in more
detail in the context of Fig.~\protect\ref{fig_densityphase}.
For comparison, circles
show the boundary between the symmetric and symmetry-broken regions
predicted by the two mode model.
Note the log scale of both axes.
}\label{fig_phasediagram}
\end{figure}

\begin{figure}
\vspace*{.2cm}
\includegraphics[angle=0,width=70mm]{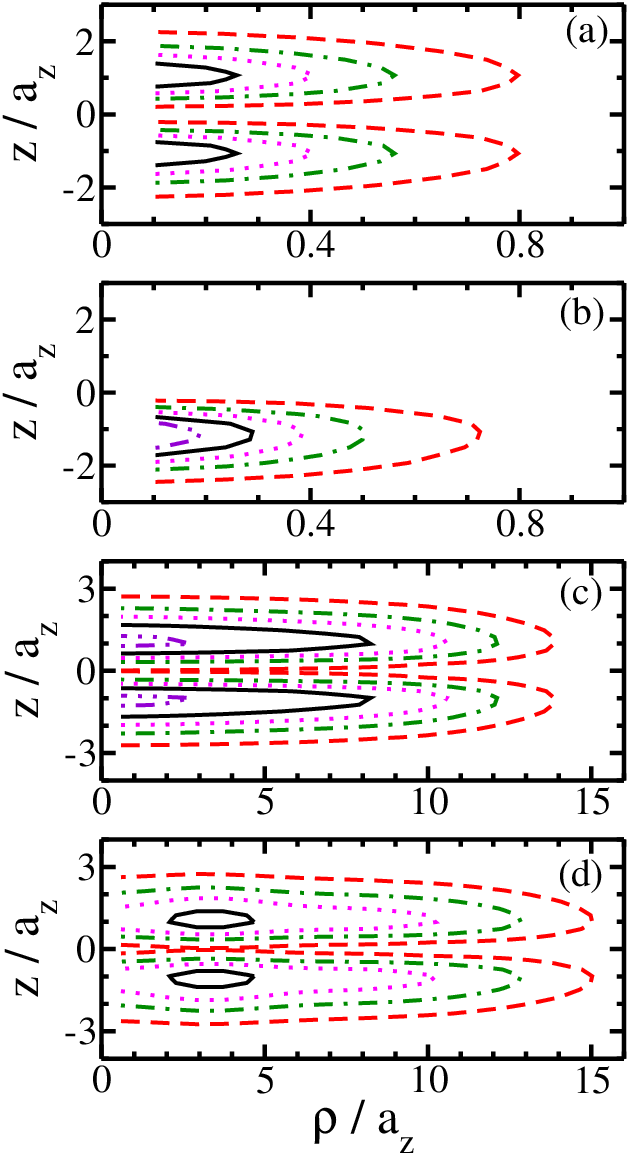}
\caption{
(Color online)
Density plots of the energetically lowest lying stationary
GP solution with $k=0$ for $b=0.2a_{z}$, $A=12 E_z$ and four
different $(D,\lambda)$ combinations:
(a) $(D,\lambda)=(0.5477,0.3)$ (symmetric solution), 
(b) $(D,\lambda)=(1.643,0.3)$ (symmetry-broken solution), 
(c) $(D,\lambda)=(316.2,10)$ (symmetric solution with density maximum 
at $\rho=0$) and 
(d) $(D,\lambda)=(474.3,10)$ (symmetric solution with density maximum
at $\rho >0$).
The contour lines are chosen equidistant in all four panels.
The dashed contours
correspond to (a) $0.05a_z^{-3}$, (b) $0.1a_z^{-3}$, (c)
$0.0001a_z^{-3}$ and (d) $0.0001 a_z^{-3}$, while
the solid contours
correspond to (a) $0.35a_z^{-3}$, (b) $0.7a_z^{-3}$, (c) 
$0.0007a_z^{-3}$ and (d) $0.0007 a_z^{-3}$.
}\label{fig_density}
\end{figure}

Figure~\ref{fig_phasediagram} summarizes the character
of the energetically lowest lying solutions with $k=0$ of the stationary
GP equation as functions of the aspect ratio $\lambda$
and the mean-field strength $D$ for 
$A=12 E_z$.
The parameter combination
$(\lambda,D)=(0.1,1)$
corresponds, e.g., 
to a Cr condensate with vanishing $s$-wave scattering length,
$\omega_z=2 \pi \times 10$Hz, $\omega_{\rho}=2 \pi \times 100$Hz 
and $N\approx 1835$.
The ``phase diagram'' consists of three regions:
Firstly,
a region where the energetically lowest lying state 
with $k=0$ of the stationary GP
equation is symmetric with respect to the $z$-axis; 
we refer to this solution 
as symmetric (``S'') throughout this paper. Examplary density profiles
are shown in Figs.~\ref{fig_density}(a), \ref{fig_density}(c) and 
\ref{fig_density}(d) 
(see below for more details).
Secondly, a 
region where the energetically lowest lying state of the stationary GP
equation is neither symmetric nor anti-symmetric
with respect to the $z$-axis; 
we refer to this solution 
as symmetry-broken (``SB'') or asymmetric.
An examplary density profile
is shown in Fig.~\ref{fig_density}(b).
And thirdly, 
a region where the stationary GP equation 
supports a high-density or collapsed solution but no gas-like
solution.
We refer to this solution 
as mechanically unstable (``U'').
Sections~\ref{sec_smalllambda} and \ref{sec_largelambda}
discuss the properties of the phase diagram in more detail
for $\lambda \lesssim 1$ and $\lambda \gtrsim 1$, respectively.

\subsection{``Small'' aspect ratio ($\lambda \lesssim 1$)}
\label{sec_smalllambda}
Figure~\ref{fig_energy}(a) shows the energy contributions 
to the total energy per particle $E_{\mathrm{tot}}/N$
for $\lambda=0.3$ and $A=12E_z$ as a function of $D$:
\begin{figure}
\vspace*{.2cm}
\includegraphics[angle=0,width=70mm]{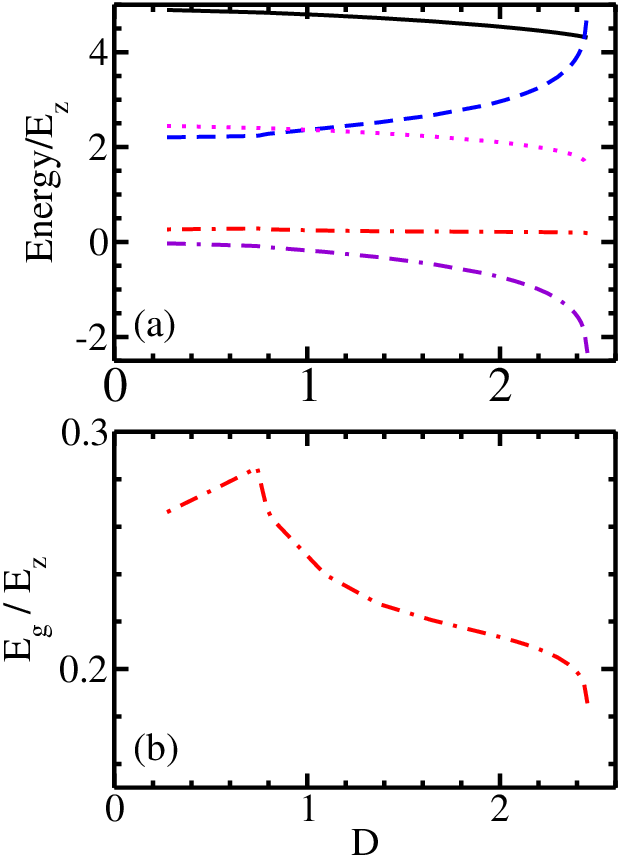}
\caption{
(Color online)
(a) Energy contributions of the energetically lowest lying stationary 
GP solution with $k=0$
as a function of $D$ for 
$\lambda=0.3$, $A = 12E_z$ and $b=0.2a_z$.
The solid, dashed, dotted, dash-dotted
and dash-dash-dotted lines show $E_{\mathrm{tot}}/N$,
$E_{\mathrm{kin}}$, $E_{\mathrm{ho}}$, $E_{\mathrm{g}}$ and $E_{\mathrm{dip}}$,
respectively.
(b) Blow-up of the Gaussian energy $E_{\mathrm{g}}$. 
$E_{\mathrm{g}}$ exhibits a kink at $D \approx 0.75$, indicating
the symmetry change (symmetric to symmetry-broken) of the energetically
lowest lying stationary GP solution.
}\label{fig_energy}
\end{figure}
the
kinetic energy per particle $E_{\mathrm{kin}}$ (dashed line);
the harmonic trap energy per particle $E_{\mathrm{ho}}$ (dotted line), 
which is defined as the expectation
value of $V_{\mathrm{ho}}$;
the Gaussian energy per particle $E_{\mathrm{g}}$ (dash-dotted line), 
which is defined as the expectation
value of $V_{\mathrm{g}}$;
and the mean-field dipole-dipole energy per particle 
$E_{\mathrm{dip}}$ (dash-dash-dotted line),
which is defined as the expectation value of the mean-field term 
[third term on the right hand side of Eq.~(\ref{eq_ham})].
A solid line shows the total energy per particle $E_{\mathrm{tot}}/N$.
The energies terminate at the critical value $D_{\mathrm{cr}}$
at which the stationary
GP equation first supports a negative energy solution.

The Gaussian energy $E_{\mathrm{g}}$ is shown on an enlarged scale in 
Fig.~\ref{fig_energy}(b). It can be seen that $E_{\mathrm{g}}$
shows a ``kink'' at $D \approx 0.75$.
We find that the other energy contributions (i.e.,
$E_{\mathrm{kin}}$, $E_{\mathrm{ho}}$ and $E_{\mathrm{dip}}$) and $E_{\mathrm{tot}}/N$ 
exhibit kinks at the 
same $D$ value. These kinks are, however, less pronounced and not 
(or hardly) visible
on the scale shown in Fig.~\ref{fig_energy}(a).
Our analysis 
shows that 
the $D$ values at which the kinks occur
coincide with the $D$ values
at which the density profiles of the energetically lowest lying
stationary GP
solutions change from symmetric 
to symmetry-broken. In most of our calculations 
for fixed $b$, $A$ and $\lambda$ but varying $D$,
we use the kink
in $E_{\mathrm{g}}$ to determine the $D$ value at which the character of the
energetically lowest lying stationary GP 
solution changes and thus to obtain
the dash-dot-dotted line in Fig.~\ref{fig_phasediagram}.
The stability of the solutions around the
symmetry to symmetry-broken transition
is discussed in Sec.~\ref{timedependent}
in the context of Figs.~\ref{fig_bdgsmalllambda} through
\ref{fig_freqjos}.

\begin{figure}
\vspace*{.2cm}
\includegraphics[angle=0,width=75mm]{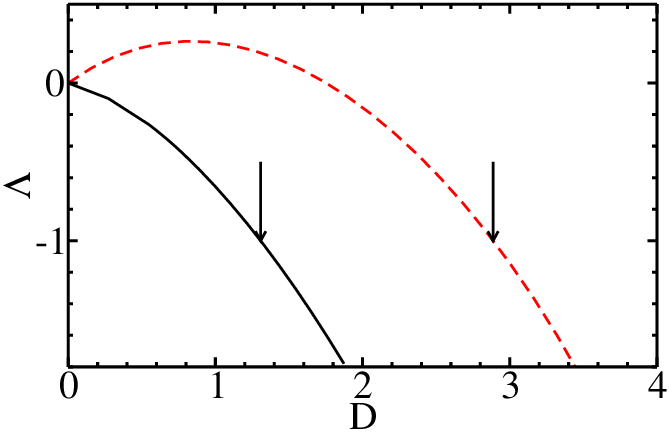}
\caption{
(Color online)
Two-mode model parameter $\Lambda$ for $A=12E_z$ and $b=0.2a_z$
as a function of $D$ for 
$\lambda=0.3$ 
(solid line)
and
$\lambda=0.4$ 
(dashed line).
Vertical arrows mark 
the $D$ values at which $|\Lambda|$ equals $1$;
for $|\Lambda| < 1$ and $>1$, the two-mode model 
predicts that the energetically lowest lying stationary 
state is symmetric and symmetry-broken, respectively.
}\label{fig_twomode}
\end{figure}

The
dash-dot-dotted lines in Fig.~\ref{fig_phasediagram}
can 
be reproduced qualitatively by
the two-mode model (see circles in Fig.~\ref{fig_phasediagram}).
To illustrate some aspects of the two-mode model,
solid and dashed lines in Fig.~\ref{fig_twomode}
show $\Lambda$ [see Eq.~(\ref{eq_biglambda})]
as a function of $D$ for $\lambda=0.3$ and $0.4$, respectively,
and $A=12E_z$ and $b=0.2a_z$.
For $|\Lambda| \le 1$ and positive $T$, the two-mode model predicts a symmetric
stationary ground state. 
For $|\Lambda|>1$, a symmetry-broken solution is supported;
if $T>0$
and $\Lambda < -1$,
the symmetry-broken state has a lower energy
than the symmetric state. 
Vertical arrows in
Fig.~\ref{fig_twomode} mark the $D$ values,
$D \approx 1.31$ and $2.89$,  at which
the transition from symmetric to
symmetry-broken occurs
for $\lambda=0.3$ 
and 
$\lambda=0.4$, respectively.
These two-mode model predictions 
(also shown as circles in Fig.~\ref{fig_phasediagram})
are slightly larger than the results obtained by solving the 
GP equation but predict
the symmetric to 
symmetry-broken transition qualitatively correctly.

It is interesting to compare the behavior of $\Lambda$,
which can be interpreted as the ratio between an effective interaction
energy and twice the tunneling energy, for $\lambda=0.3$ 
and $0.4$ (solid and dashed lines in Fig.~\ref{fig_twomode}).
For $\lambda=0.3$, an increase of the mean-field strength $D$ leads to a 
monotonic decrease of $\Lambda$.
For $\lambda=0.4$, in contrast, $\Lambda$ first increases, reaches a maximum at
$D \approx 0.87$ and then decreases monotonically.
We find that the onsite energy $U$ and the offsite energy $B$ are both
negative for all $D$ shown in Fig.~\ref{fig_twomode}. 
A change of the aspect ratio $\lambda$ effectively changes the 
strength of the dipole-dipole interaction, leading to a less 
attractive $U$ than $B$, and thus to a positive $\Lambda$, 
for small $D$ and $\lambda=0.4$.
For $\lambda=0.3$, in contrast, the onsite energy $U$ is always more
negative than the offsite energy $B$, resulting in a negative 
$\Lambda$ for all $D$.

In addition to the barrier height $A=12E_z$, 
we considered smaller barrier heights $A$, 
in particular $A=4E_z$ and
$8E_z$,
for a few selected
$\lambda$ values.
Our calculations
suggest that the dash-dot-dotted line in Fig.~\ref{fig_phasediagram}
(i.e., the line that marks the symmetric to symmetry-broken transition)
moves to larger $D$ values with decreasing $A$ while the dash-dotted line
(i.e., the line that marks the symmetry-broken 
to unstable transition)
remains approximately unchanged with decreasing $A$.
The dependence of the dash-dot-dotted line on $A$ for fixed 
$\lambda$ and $b$ can be explained by applying the two-mode model.
As $A$ decreases, the tunneling energy $T$ becomes more important compared
to the absolute value of
the effective interaction energy $U-B$. This implies that
$|\Lambda|$ decreases with decreasing $A$ (for fixed $\lambda$ and $b$).
Correspondingly, 
a larger $D$ is required
for the two-mode model 
condition $|\Lambda|=1$, which signals the symmetric to
symmetry-broken transition, to be fulfilled. 
The fact that the dash-dotted line in Fig.~\ref{fig_phasediagram}
remains to first order unchanged with decreasing $A$ 
is due to the fact that the density of the system
prior to collapse is located predominantly in one of the wells.
This implies that the density prior to
collapse is only weakly dependent on $A$, thus explaining the comparatively
small dependence of the 
dash-dotted line on $A$ for the parameter combinations
investigated.

We note at this point that the linear stationary Schr\"odinger equation
permits only symmetric and anti-symmetric solutions but no 
symmetry-broken solutions. 
This fact emphasizes that the transition from symmetric 
to symmetry-broken is driven by mean-field interactions. 
Furthermore, this fact implies that the symmetry-broken solution should
disappear if sufficiently many higher order corrections 
to the mean-field GP equation are taken into account
(see, e.g., Ref.~\cite{ragh99}).
In this sense, the appearance of the symmetry-broken 
region in the phase diagram is
an artefact of the mean-field formalism. It is,
however, intimately related to
the dynamical phenomena of Josephson oscillation and 
macroscopic quantum self-trapping, both
of which have been observed experimentally for $s$-wave 
interacting BECs. 
We return to these considerations
in Sec.~\ref{timedependent} in the context 
of the discussion of Figs.~\ref{fig_bdgsmalllambda} through
\ref{fig_freqjos}.

\subsection{``Large'' aspect ratio ($\lambda \gtrsim 1$)}
\label{sec_largelambda}
\begin{figure}
\vspace*{.2cm}
\includegraphics[angle=0,width=75mm]{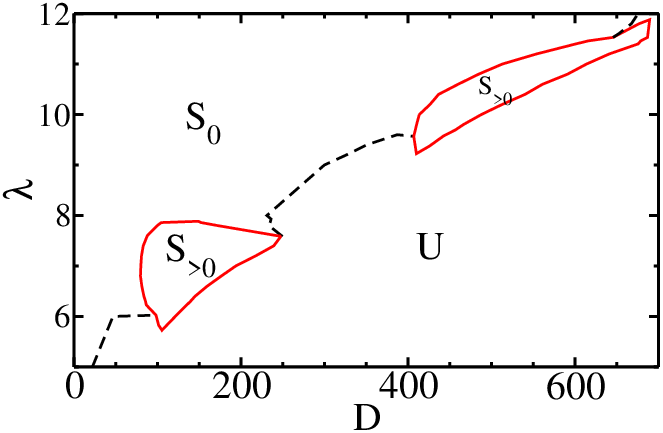}
\caption{
(Color online)
Blow-up  (with more detail)
of the
phase diagram for $b=0.2a_{z}$ and $A = 12 E_z$ 
shown in Fig.~\protect\ref{fig_phasediagram}:
The region where the energetically lowest lying 
symmetric stationary GP solution 
has its density maximum at $\rho=0$ is labeled by ``S$_0$''
and that
where the energetically lowest lying symmetric stationary GP solution 
has its density maximum at $\rho>0$ by ``S$_{>0}$''.
}\label{fig_densityphase}
\end{figure}

Figure~\ref{fig_densityphase} shows
an enlargement of the large $\lambda$ region of Fig.~\ref{fig_phasediagram}
using a linear scale
for both $\lambda$ and $D$. 
The S$_0$ region of the phase diagram is characterized
by GP solutions
whose density maxima 
are located at $\rho=0$ [see Figs.~\ref{fig_density}(a) and 
\ref{fig_density}(c)
for examples] 
while the S$_{>0}$ region of the phase diagram is charactericed by GP
solutions whose density maxima are 
located at $\rho >0$ [see Fig.~\ref{fig_density}(d)
for an example]. 
The latter class of density profiles
only exists in a narrow parameter region of the phase diagram;
in particular, these solutions only arise
for pancake-shaped confining potentials and not
for cigar-shaped confining potentials. Furthermore, the solutions
with S$_{>0}$ character are unique to dipolar gases, i.e.,
they are not observed for purely $s$-wave interacting gases, and 
thus
directly reflect the anisostropic long-range nature
of the dipole-dipole interactions.

The two different classes of symmetric solutions have
previously been characterized for $A=0$, i.e., for a pancake-shaped
trapping geometry without barrier~\cite{rone07,wils08}. 
In those studies, a dipolar BEC with density maximum
at $\rho>0$ was termed ``red blood cell'', as its isodensity
surface is reminiscent of the shape of a red blood cell.
The S$_{>0}$ regions in Fig.~\ref{fig_densityphase} are characterized by 
the formation of two staggered red blood cells. 
Section~\ref{timedependent} shows that
the dynamical instability near $D_{\mathrm{cr}}$ 
of the stationary $k=0$ ground state
solutions of types
S$_0$ and S$_{>0}$ is distinctly different.

Figure~\ref{fig_densityphase} shows that, generally speaking,
the $D$ value at which the
dipolar gas becomes unstable increases with increasing $\lambda$.
This
trend
can be understood by realizing that an increase of $\lambda$ leads to 
a ``flattening'' of the system so that the dipoles
interact effectively more repulsively. 
The boundary near the stable and unstable regions shows a rich structure:
i) As already noted
above, S$_{>0}$ islands in which the density profiles are structured exist.
ii)
The boundary between the S$_0$ and the U regions of the phase
diagram changes non-monotonically. For $D\approx 240$, e.g., the system is
mechanically stable for $\lambda \gtrsim 8.13$, mechanically
unstable for $8.13 \gtrsim \lambda  \gtrsim 7.72$, 
and then again mechanically stable
for a small $\lambda$ regime ($7.72 \gtrsim \lambda \gtrsim 7.42$).

\begin{figure}
\vspace*{.2cm}
\includegraphics[angle=0,width=75mm]{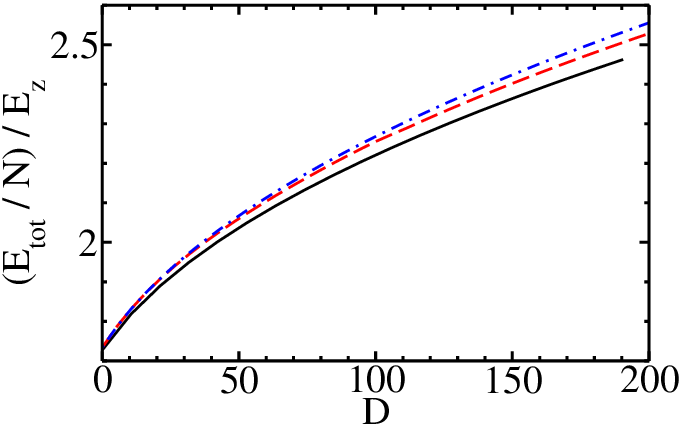}
\caption{
(Color online)
A solid line shows the total energy per particle $E_{tot}/N$
of the energetically lowest lying stationary
GP solution with $k=0$
as a function of $D$ for
$\lambda=7$, $A = 12E_z$ and $b=0.2a_z$ obtained numerically.
For comparison, dashed and dash-dotted lines show $E_{tot}/N$
obtained using the variational
six- and four-parameter
variational wave functions (see text for details).
The density of the system changes from S$_0$ to S$_{>0}$ character 
at $D\approx 35$ and $80.03$ 
for the six-parameter variational and the full
numerical calculations, respectively.
}\label{fig_energyvar}
\end{figure}

We find that some, though not all, of the features of the 
phase diagram can be reproduced qualitatively by
a simple variational wave function $\psi_{var}(\rho,z)$,
\begin{eqnarray}
\label{eq_var}
\psi_{var}(\rho,z)=
\left[\exp\left( - \frac{\rho^2}{2b_1^2}\right) + b_2 
 \exp\left( - \frac{\rho^2}{2b_3^2}\right) 
\right] 
\times \nonumber \\
\left[ 
\exp\left( - \frac{z^2}{2b_4^2}\right) + 
b_5 \left( 1- \frac{z^2}{\lambda a_z^2} \right)
\exp\left( - \frac{z^2}{2b_6^2}\right)
\right],
\end{eqnarray} 
where $b_1-b_6$ denote variational parameters
that are optimized by minimizing the energy per particle.
For $b_2=b_5=0$, $\psi_{var}$ reduces to the commonly used
variational wave function of purely Gaussian shape.
The second term in the first square bracket on the right hand side of 
Eq.~(\ref{eq_var})
has been added to allow for the description of 
densities of S$_{>0}$ character while
the second term in the second square bracket on the right hand side of 
Eq.~(\ref{eq_var}) has been added
to account for the 
Gaussian barrier along the $z$-direction.
Figure~\ref{fig_energyvar} 
compares the total energy per particle from our 
variational calculation (dashed line)
with that from the full numerical calculation (solid line)
for $A=12E_z$, $\lambda=7$ and $b=0.2a_z$.
The variational energy is less than 2~\% higher than the 
energy obtained from the full
numerical calculation.
We find that the density of
the dipolar gas
changes from S$_0$ to S$_{>0}$ character at $D\approx 35$,
compared to
$D=80.03$ obtained from the full calculation. 
For both sets of 
calculations, the energy and its derivative change smoothly as the 
system undergoes the structural change 
from S$_0$ to S$_{>0}$.
For comparison, a dash-dotted line in Fig.~\ref{fig_energyvar}
shows the
energy per particle for $\psi_{var}$ with $b_2=0$
(we refer to this variational wave function as four-parameter wave function), 
i.e., for a 
wave function that is not sufficiently flexible to describe structured
ground state densities of red blood cell shape.
As expected, this variational 
wave function results in somewhat higher energies.

The variational wave function 
$\psi_{var}$ predicts S$_0$ to S$_{>0}$ transitions for all
aspect ratios $\lambda$ between
$5$ and $12$, indicating that
it is not flexible enough to describe the island character
of the S$_{>0}$ regions of the phase diagram
and, furthermore, that the S$_0$ to S$_{>0}$ transition is
driven by a delicate balance between the different energy contributions.
Motivated by calculations presented in
Ref.~\cite{rone07}, we expect that the variational
four-parameter wave function can qualitatively
reproduce the existence of alternating stable and unstable
regions of the phase diagram as 
$\lambda$ is changed for fixed
$D$ and $A$ (see our discussion above for $D\approx240$ and $A=12E_z$);
we have, however, not checked this explicitly.
Lastly, we note that the variational six-parameter wave function
predicts a stable
dipolar
gas even for fairly large $D$ (i.e., $D$ values larger than
those shown in
Fig.~\ref{fig_energyvar}) while the full numerical calculation
predicts collapse at $D \approx 190.49$.

\section{Discussion of dynamical studies}
\label{timedependent}
This section presents Bogoliubov de Gennes excitation spectra and 
discusses the corresponding 
eigenmodes.
For small $\lambda$ 
and appropriate $D$ (see Sec.~\ref{sec_smalllambda2}),
the lowest non-vanishing Bogoliubov de Gennes
excitation frequency is identified as the Josephson oscillation
frequency $\omega_{\mathrm{J}}$. Comparisons with results obtained
by time-evolving a properly prepared initial state 
and by applying the 
two-mode model equations are presented.
In the 
regime where the symmetric solutions are of types 
$S_0$ and $S_{>0}$,
respectively  (large $\lambda$, see Sec.~\ref{sec_largelambda2}), 
the decay mechanisms are identified.
Furthermore, the character of
various (avoided) 
crossings of the
excitation frequencies
is revealed.

\subsection{``Small'' aspect ratio ($\lambda \lesssim 1$)}
\label{sec_smalllambda2}
\begin{figure}
\vspace*{.2cm}
\includegraphics[angle=0,width=75mm]{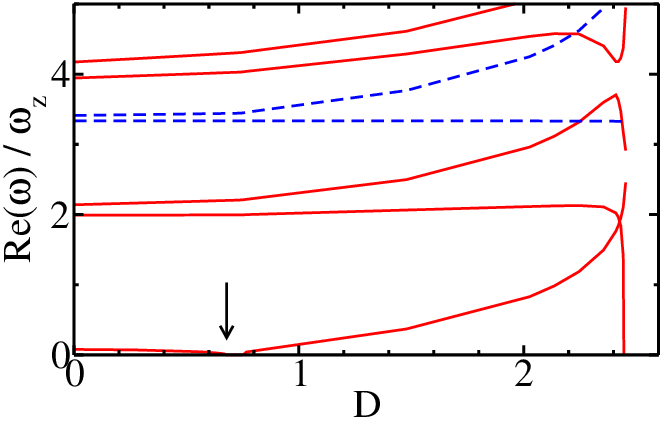}
\caption{
(Color online)
Excitation spectrum 
obtained by
solving the Bogoliubov de Gennes equations 
as a function of $D$ 
for $A=12E_z$, $b=0.2 a_{z}$ and $\lambda=0.3$.
The real parts of the frequencies for $k=0$ and 1 are shown by
solid and dashed lines, respectively.
The vertical arrow indicates the $D$ value, $D\approx 0.68$,
at which the real part of the lowest non-vanishing $k=0$ Bogoliubov
de Gennes frequency
vanishes.
}\label{fig_bdgsmalllambda}
\end{figure}

Figure~\ref{fig_bdgsmalllambda} shows the excitation spectrum 
as a function of $D$
obtained by solving the Bogoliubov de Gennes equations for $A=12E_z$,
$b=0.2a_{z}$ and $\lambda=0.3$.
The spectrum is characterized by three distinct features that will
be elaborated on in the following paragraphs:
i) The real part of the lowest $k=0$ frequency vanishes  at 
$D \approx 0.68$,
and ``reappears'' at $D \approx 0.75$.
ii) The $k=0$ frequencies show a series of crossings (or
avoided crossings) at $D \approx 2.42$.
iii) At slightly larger $D$ values,
i.e., near $D \approx 2.45$,
the real part of several $k=0$ excitation frequencies vanishes.

\begin{figure}
\vspace*{.2cm}
\includegraphics[angle=0,width=70mm]{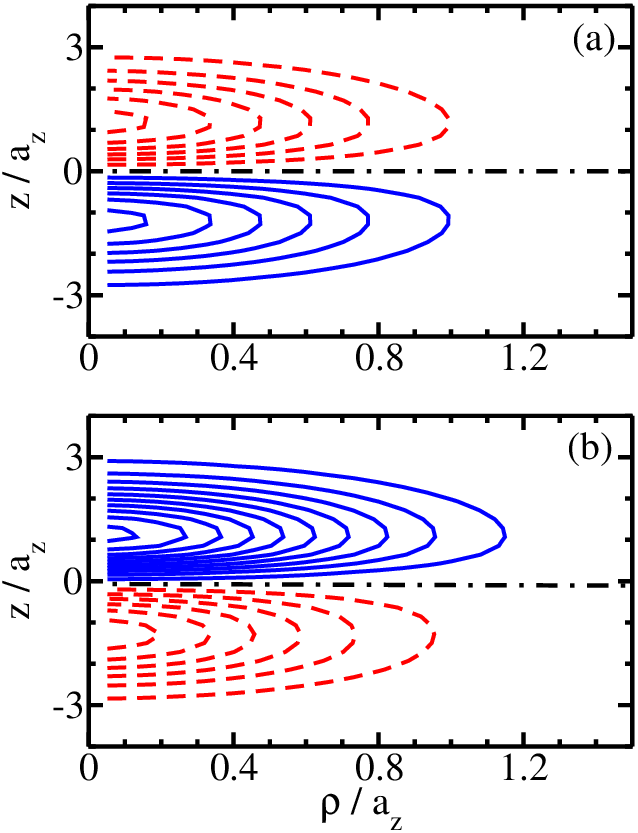}
\caption{
(Color online)
Bogoliubov de Gennes eigenmodes $\bar{f}(\rho,z)$ corresponding to the lowest
non-vanishing $k=0$ 
frequency for $\lambda=0.3$,
$A=12E_z$, $b=0.2 a_{z}$ and (a) $D=0.6573$ and (b) $D=0.7668$. 
The contours are chosen equidistant, with solid
and dashed lines corresponding
to positive and negative values
of $\bar{f}$.
The dash-dotted lines indicate the nodal lines of 
$\bar{f}$.
}\label{fig_eigenmode1}
\end{figure}

We first  discuss the regime i) 
around $D \approx 0.68-0.75$.
Figure~\ref{fig_eigenmode1}(a) shows the 
Bogoliubov de Gennes eigenmode 
$\bar{f}(\rho,z)$
that
corresponds to the lowest 
non-vanishing $k=0$ frequency
for $D=0.6573$, $A=12 E_z$, $\lambda=0.3$ and $b=0.2 a_{z}$.
For these parameters, the energetically lowest lying
stationary GP solution is symmetric
and the corresponding eigenfrequency has a finite real part and 
vanishing
imaginary part (see Fig.~\ref{fig_bdgsmalllambda}).
Since Bogoliubov de Gennes functions with $k=0$
have no explicit $\varphi$ dependence,
the eigenmode shown in Fig.~\ref{fig_eigenmode1}(a) 
corresponds to a situation where the population oscillates
with frequency $\omega$ between the left
and the right well as a function of time. 
The lowest non-vanishing $k=0$ 
frequency can thus be identified
as the Josephson oscillation frequency $\omega_{\mathrm{J}}$
(see also below).
For comparison, 
Fig.~\ref{fig_eigenmode1}(b) shows the 
Bogoliubov de Gennes eigenmode
$\bar{f}(\rho,z)$
corresponding to
the lowest non-vanishing $k=0$ frequency
for 
$D=0.7668$ (i.e., in the regime where 
the frequency has ``reappeared'' and where the energetically lowest
lying stationary GP solution with $k=0$ is symmetry-broken) 
and the same $A$, $E_z$ and $b$ values as before.
In this case, the
asymmetry of the eigenmode indicates 
that there is population transfer between 
the left and the right wells but that there is, on average, more population 
in the right than in the left well.
This behavior is identified as macroscopic quantum self-trapping.
Our interpretation
of the Bogoliubov de Genne
eigenmodes is supported by our time-dependent calculations.

\begin{figure}
\vspace*{.2cm}
\includegraphics[angle=0,width=70mm]{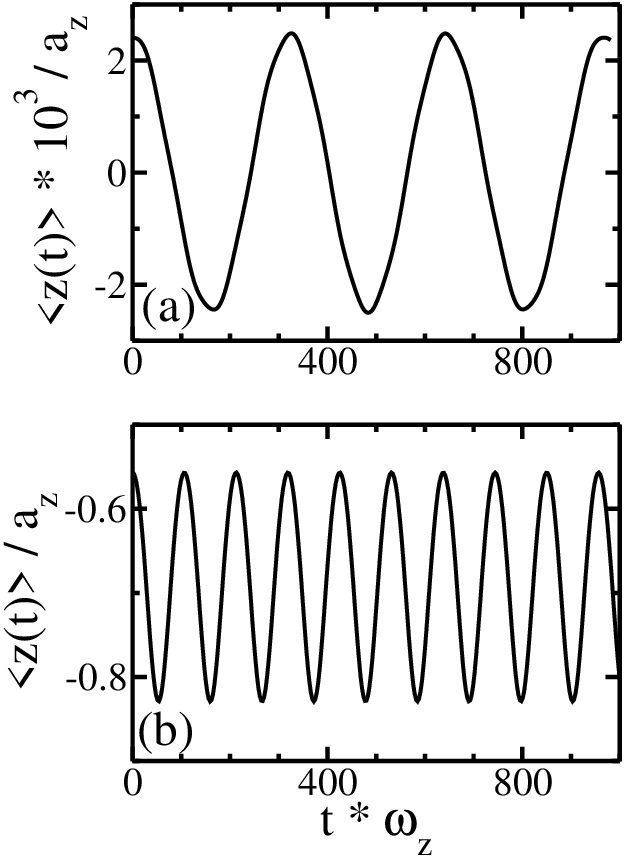}
\caption{
(Color online)
Solid lines show 
the expectation value 
$\langle z(t) \rangle$
(see text)---calculated 
by time evolving a given initial state
according to
the mean-field Hamiltonian, Eq.~(\protect\ref{eq_ham})---as 
a function of time $t$ for 
$A=12 E_z$,
$\lambda=0.3$, 
$b=0.2a_z$ and (a) $D=0.6573$ (Josephson oscillation regime) and 
(b) $D=0.7668$ (macroscopic quantum self-trapping regime).
In panel (a), the initial state is prepared according to
Eq.~(\ref{eq_psitwomode})
with $Z(0)=0.002$ and $\phi(0)=0$.
In panel (b), the initial state is prepared by adding a small
amount of random noise to the energetically lowest
lying stationary GP solution.
}\label{fig_oscjos}
\end{figure}

In our time-dependent studies near the 
symmetric to symmetry-broken transition,
we prepare
an initial state 
and time evolve
it according to the mean-field Hamiltonian $H$, 
Eq.~(\ref{eq_ham}).
As for $s$-wave interacting BECs, the system dynamics
can be divided into two categories (see also above):
A regime where the population is transferred back and forth
between the left well and the right well (this is the Josephson
oscillation regime) and a regime
where the time averaged
population is asymmetrically divided among the two wells
(this is the macroscopic quantum self-trapping regime).
Figures~\ref{fig_oscjos}(a) and \ref{fig_oscjos}(b) 
show the time evolution of
the expectation 
value
$\langle z(t) \rangle$
for $D=0.6573$ and 
$D=0.7668$, respectively.
Here, $\langle z(t) \rangle$ is obtained by 
calculating the expectation value of $z$ with respect to the GP density
at each time step. The expectation value $\langle z(t) \rangle$
is related to but not identical to the population
difference $Z(t)$ introduced in Sec.~\ref{twomode}.
For these $D$ values, the energetically lowest lying stationary GP solution
is symmetric and symmetry-broken, respectively.
For $D=0.6573$, the initial state 
is prepared
according to Eq.~(\ref{eq_psitwomode})
with 
$\phi(0)=0$ and $Z(0)=0.002$. Figure~\ref{fig_oscjos}(a) shows that
$\langle z(t) \rangle$ oscillates between positive and negative values of 
equal magnitude and 
that the time average of $\langle z(t) \rangle$ 
over a period gives zero.
For $D=0.7668$, the initial state is prepared by adding a small amount
of random noise to the energetically lowest lying stationary GP 
solution.
Figure~\ref{fig_oscjos}(b) shows that
$\langle z(t) \rangle$ oscillates about a negative
value and
that the time average of $\langle z(t) \rangle$ gives a non-zero value.
An  analysis of the time evolution of the
density profiles confirms that the system is in
the Josephson regime and in the macroscopic 
quantum self-trapping regime, respectively. 

\begin{figure}
\vspace*{.2cm}
\includegraphics[angle=0,width=75mm]{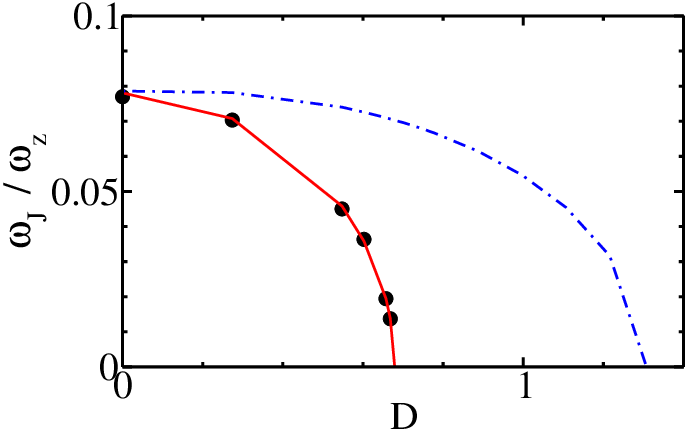}
\caption{
(Color online)
Josephson oscillation frquency $\omega_{\mathrm{J}}$ as a function of
$D$ for $\lambda=0.3$, $A=12E_z$ and $b=0.2 a_{z}$.
The circles show the
Josephson oscillation frequency $\omega_{\mathrm{J}}$ 
obtained from our time-dependent study, in which the 
initial state is prepared according to Eq.~(\ref{eq_psitwomode})
with $Z(0)=0.002$ and $\phi(0)=0$
and then time evolved according to the mean-field
Hamiltonian $H$, Eq.~(\ref{eq_ham}).
The solid line shows the lowest non-vanishing $k=0$ Bogoliubov de Gennes
excitation frequency.
For comparison,
a dash-dotted line shows the two-mode model prediction 
$\omega_{\mathrm{J,TM}}$.
}\label{fig_freqjos}
\end{figure}

To determine the oscillation frequency from the time evolution
of $\langle z(t) \rangle$, 
we Fourier transform $\langle z(t) \rangle$
and record the center of the 
dominant peak for various parameter combinations.
Circles in Fig.~\ref{fig_freqjos}
show the resulting Josephson oscillation frequency 
$\omega_{\mathrm{J}}$ for $A=12E_z$,
$\lambda=0.3$ and $b = 0.2a_{z}$. 
The agreement between the frequency obtained from the Fourier analysis
(circles in Fig.~\ref{fig_freqjos}) 
and the lowest non-vanishing $k=0$ Bogoliubov de Gennes
excitation frequency (solid line in Fig.~\ref{fig_freqjos})
is excellent.
The $D$ value at which the
Josephson oscillation frequency obtained by Fourier transforming
$\langle z(t) \rangle$ vanishes, coincides, within our numerical accuracy,
with that at which the lowest non-vanishing
$k=0$ Bogoliubov de Gennes excitation frequency 
becomes imaginary.
Notably, this $D$ value, $D\approx 0.68$,
is slightly smaller than the $D$
value at which the energetically lowest lying stationary GP solution
changes from symmetric to symmetry-broken ($D \approx 0.75$).

We find that 
the 
lowest non-vanishing $k=0$ Bogoliubov de Gennes frequency 
for $D=0.7668$ and $\lambda=0.3$ is about 30\% smaller than 
the oscillation frequency extracted from Fig.~\ref{fig_oscjos}(b),
i.e., $\omega=0.059 \omega_z$.
The fact that the Bogoliubov
de Gennes excitation frequency 
differs notably from 
the frequency obtained by 
Fourier-transforming $\langle z(t) \rangle$
might be due to the 
approximate nature of the Bogoliubov de Gennes equations.

For comparison, a dash-dotted line in Fig.~\ref{fig_freqjos}
shows the Josephson 
oscillation frequency $\omega_{\mathrm{J,TM}}$,
Eq.~(\ref{eq_omegaj}),  predicted by the two-mode
model. Figure~\ref{fig_freqjos}
shows that the two-mode model provides a qualitatively 
but not quantitatively correct description
of the Josephson oscillation frequency.
The fact that the two-mode model
does not allow for 
quantitative predictions for all $D$ is likely due to the fact that 
the modes $\Phi_{\mathrm{L}}$ and $\Phi_{\mathrm{R}}$ 
are not  entirely located in the left 
well and in the
right well, respectively, but that the left mode ``leaks'' into the 
right well and the right mode into the left 
well. This has been discussed in some detail
in Ref.~\cite{xion09}, which employs a slightly modified version of the 
two-mode model.
In an attempt to obtain a better simple quantitative
description of the system dynamics, we applied the 
improved two-mode model proposed in Ref.~\cite{anan06}.
For the cases considered, we find that this model leads only to small 
changes compared to the simple two-mode
model
applied
above and does not provide a significantly improved description.
In the future, it may be interesting to apply a multi-mode model.

\begin{figure}
\vspace*{.2cm}
\includegraphics[angle=0,width=120mm]{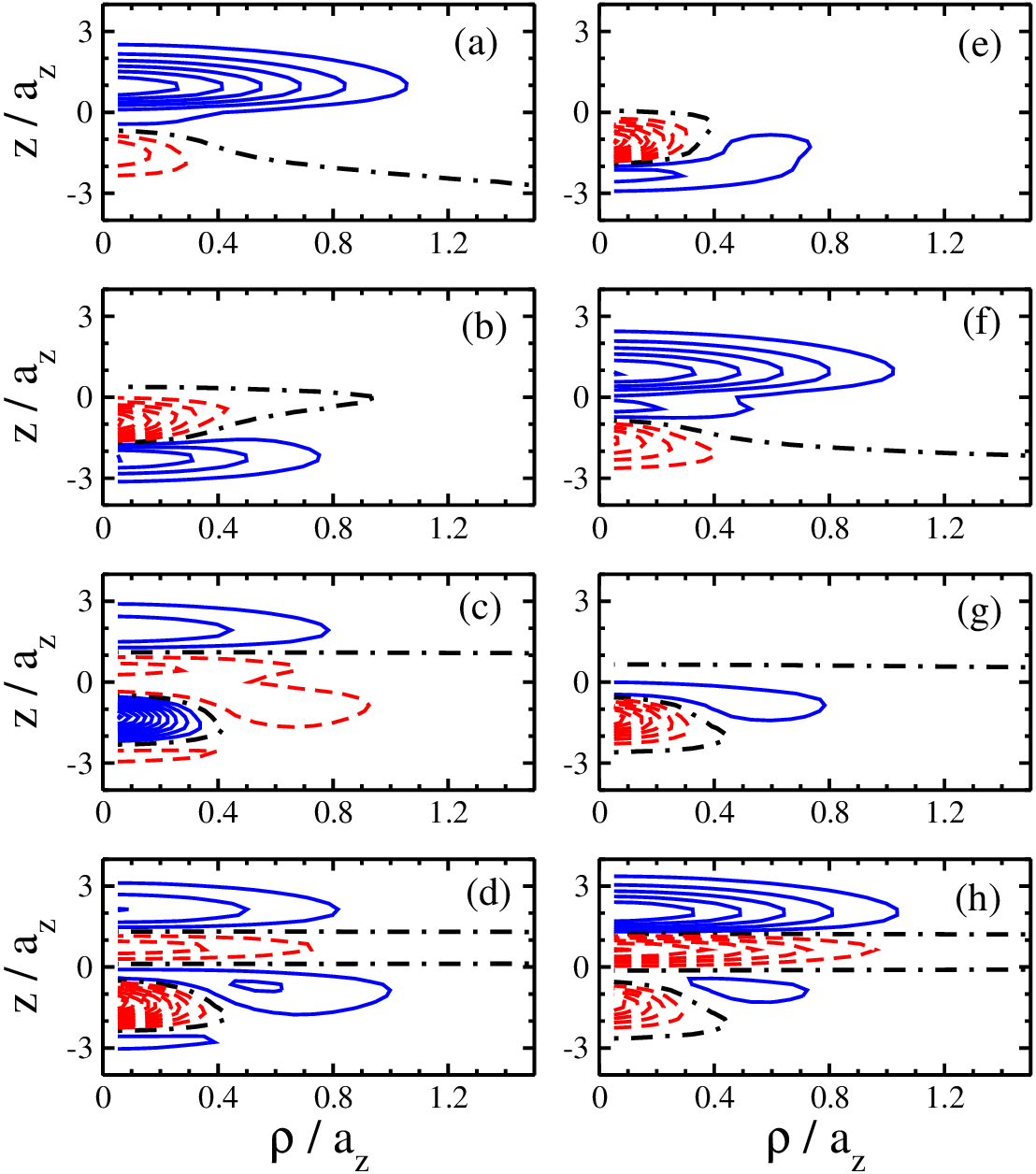}
\caption{
(Color online)
Bogoliubov de Gennes eigenmodes $\bar{f}(\rho,z)$ corresponding to the four
lowest non-vanishing $k=0$ frequencies for $\lambda=0.3$,
$A=12E_z$, $b=0.2 a_{z}$.
Panels (a)-(d) show 
$\bar{f}$
corresponding to the lowest, second lowest, third lowest and fourth 
lowest frequencies for $D=2.410$ 
(i.e., just before the crossing) 
while
panels (e)-(h) show 
$\bar{f}$
corresponding to the lowest, second lowest, third lowest and fourth 
lowest frequencies for $D=2.443$ 
(i.e., just after the crossing).
The contours are chosen equidistant, with solid and dashed
lines corresponding respectively
to positive and negative values
of 
$\bar{f}$.
}\label{fig_eigenmode2}
\end{figure}

We now discuss the regime ii) near $D\approx 2.42$, where the $k=0$
frequencies show (avoided) crossings.
To shed light on these (avoided) crossings,
Figs.~\ref{fig_eigenmode2}(a)-(d) show the 
Bogoliubov de Gennes
eigenmodes 
$\bar{f}(\rho,z)$
corresponding to 
the four lowest non-vanishing $k=0$ frequencies just before the 
crossing (i.e., for $D=2.410$), while 
Figs.~\ref{fig_eigenmode2}(e)-(h) show those
corresponding to the four lowest $k=0$ frequencies just after the 
crossing (i.e., for $D=2.443$).
Dash-dotted lines in Fig.~\ref{fig_eigenmode2} indicate the 
nodal lines of the Bogoliubov de Gennes eigenmodes.
While some of these nodal lines are to first order only dependent on
$z$, others depend in a non-trivial manner on $\rho$ and $z$.
In the following we discuss a few key features of the
eigenmodes shown in Fig.~\ref{fig_eigenmode2}.
The eigenmode corresponding to the
lowest frequency 
extends over both wells just before the
crossing [see Fig.~\ref{fig_eigenmode2}(a)] 
and is located predominantly in one of the wells 
just after the crossing [see Fig.~\ref{fig_eigenmode2}(e)].
The eigenmode corresponding to the
second lowest non-vanishing frequency, in turn, 
is located predominantly in one of the wells 
just before the crossing [see Fig.~\ref{fig_eigenmode2}(b)]
and extends over both wells just after the
crossing [see Fig.~\ref{fig_eigenmode2}(f)].
The third and fourth lowest excitation frequencies show an
avoided crossing at $D\approx 2.41$ (see Fig.~\ref{fig_bdgsmalllambda}).
The eigenmode corresponding to the third lowest
non-vanishing $k=0$ frequency 
has a fairly small amplitude in the right well
before the avoided crossing [see Fig.~\ref{fig_eigenmode2}(c)];
after the avoided crossing, 
the eigenmode corresponding to the third lowest
$k=0$ frequency is located essentially entirely in the left well
[see Fig.~\ref{fig_eigenmode2}(g)].
The eigenmode corresponding to the forth lowest frequency, in turn,
changes comparatively little as $D$ increases
[see Figs.~\ref{fig_eigenmode2}(d)
and \ref{fig_eigenmode2}(h)].

Next, we turn to regime iii) near $D \approx 2.45$, where
the real part of several $k=0$ Bogoliubov de Gennes 
excitation frequencies vanishes.
The softening of these frequencies is inherently related to
the transition from the stable symmetry-broken region
to the unstable region of the phase diagram.
The Bogoliubov excitation spectrum 
for $A_z=12E_z$, $\lambda=0.3$ and $b=0.2a_{z}$ shows that the 
real part of the lowest
$k=0$ frequency vanishes at $D \approx 2.45$, which
is slightly smaller than the $D$ value at which the stationary GP
equation starts supporting unbounded negative energy solutions.
In particular, for the parameter combination considered, the difference is
about 0.6\%.
The Bogoliubov de Gennes excitation spectrum indicates that the collapse
is triggered by the lowest $k=0$ mode. The 
corresponding eigenmode [see Fig.~\ref{fig_eigenmode2}(e)] 
shows that, as might be expected naively,
the density grows appreciably in the well that supports the
majority of  the
population. This interpretation is supported by
our time-dependent studies.
Following the lowest $k=0$ mode, 
the eigenmode corresponding to the third lowest non-vanishing
$k=0$ frequency becomes soft. As indicated in Fig.~\ref{fig_eigenmode2}(g),
this eigenmode is, not unexpectedly, 
also located predominately in one of the wells.
To summarize, the collapse of the system can be characterized
as a global collapse in which the density maximum increases in 
one of the wells, with the density maximum 
being located at $\rho=0$.

\subsection{``Large'' aspect ratio ($\lambda \gtrsim 1$)}
\label{sec_largelambda2}
This section discusses 
selected Bogoliubov de Gennes excitation spectra for larger 
$\lambda$. In particular, we focus on $\lambda=5$ and $\lambda=7$,
for which the energetically lowest lying stationary 
GP solution prior to collapse is of type $S_0$ and $S_{>0}$, respectively.

\begin{figure}
\vspace*{.2cm}
\includegraphics[angle=0,width=70mm]{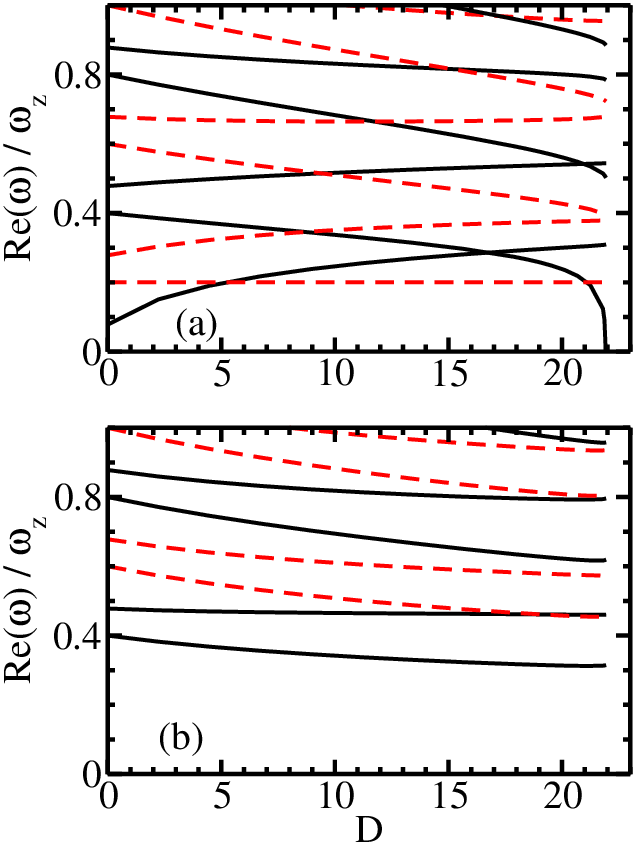}
\caption{
(Color online)
Excitation spectrum 
obtained by
solving the Bogoliubov de Gennes equations 
as a function of $D$ 
for $A=12E_z$, $b=0.2 a_{z}$ and $\lambda=5$.
Solid and dashed lines in panel~(a) show the
real parts of the Bogoliubov
de Gennes excitation frequencies for $k=0$ and $1$,
while solid and dashed lines
in panel~(b) show those
for $k=2$ and $3$.
}\label{fig_bdglargelambda1}
\end{figure}

\begin{figure}
\vspace*{.2cm}
\includegraphics[angle=0,width=70mm]{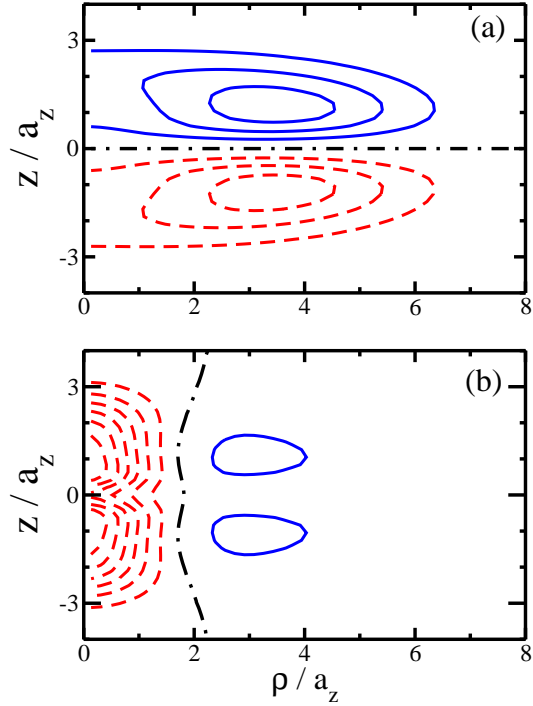}
\caption{(Color online)
Bogoliubov de Gennes eigenmodes $\bar{f}(\rho,z)$ corresponding to the lowest
non-vanishing $k=0$ 
frequency for $\lambda=5$,
$A=12E_z$, $b=0.2 a_{z}$ and (a) $D=13.42$ 
and (b) $D=21.87$.
The contours are chosen equidistant, with solid
and dashed lines corresponding
to positive and negative values
of 
$\bar{f}$.
The dash-dotted lines indicate the nodal line of 
$\bar{f}$.
}\label{fig_eigenmode3}
\end{figure}

Figure~\ref{fig_bdglargelambda1} shows the 
excitation spectrum 
as a function of $D$ 
for $\lambda=5$, $A=12E_z$ and $b=0.2a_z$.
The two lowest non-vanishing $k=0$ frequencies 
[solid lines in Fig.~\ref{fig_bdglargelambda1}(a)] cross at $D \approx17$.
Prior to the crossing, the lowest $k=0$ 
frequency corresponds to an eigenmode whose nodal line 
is parameterized by $z=0$ 
[see dash-dotted line in Fig.~\ref{fig_eigenmode3}(a)].
Correspondingly, the eigen mode describes density oscillations between the 
left and the right well with, on average, equal densities 
in each of the two wells.
For $D \approx 17-22$ (i.e., after the crossing), the
nodal line of the lowest non-vanishing $k=0$ eigenmode is 
to a good approximation independent of $z$ and can be parametrized by 
$\rho \approx 2 a_z$ to $\approx 3a_z$.
As an example,
Fig.~\ref{fig_eigenmode3}(b) shows the 
eigenmode for $D= 21.87$, which is
just slightly smaller than the $D$ value
at which the real part of the corresponding Bogoliubov de
Gennes frequency vanishes. 
In particular, the lowest Bogoliubov de Gennes mode becomes soft at
$D \approx 21.92$ while the stationary GP equation ceases
to a support a positive energy solution with $k=0$ at 
a somewhat larger $D$ value, namely at $D \approx 22.03$.

Figures~\ref{fig_bdglargelambda1} and \ref{fig_eigenmode3}(b)
suggest that the collapse of the system is 
triggered by the lowest $k=0$ mode and that the collapse
is associated with an increase of the peak density 
at $\rho=0$ in each of the two wells. 
The collapse can thus be characterized 
as a local collapse as opposed to a global collapse. 
The local nature of the collapse
(i.e., the fact that the peak density grows simultaneously in two
distinct regions)
can be traced back directly to the 
presence of the comparatively large Gaussian barrier.
The chemical potential $\mu$ takes values 
around $2E_z$ for the parameter range considered
in Fig.~\ref{fig_bdglargelambda1} and is thus significantly
smaller than the barrier height $A$, $A=12E_z$.
In the limit that the Gaussian barrier vanishes~\cite{rone07}, the 
collapse becomes global.
In this case,
a similar nodal pattern of the eigenmode 
was found (see Fig.~2Ic of Ref.~\cite{rone07}; 
the larger number of nodal lines in this plot
can be traced back to the larger $\lambda$ value)
and the collapse is associated with a radial roton.

\begin{figure}
\vspace*{.2cm}
\includegraphics[angle=0,width=70mm]{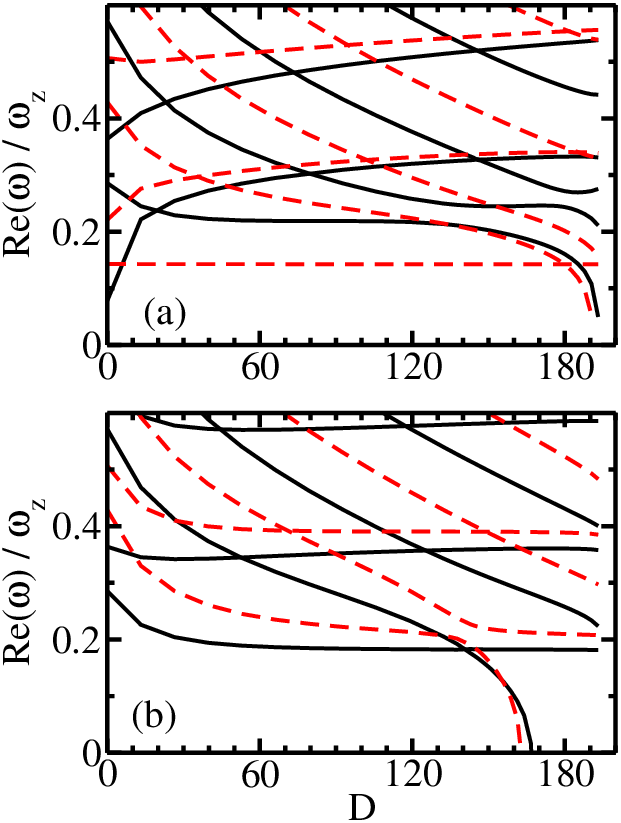}
\caption{
(Color online)
Excitation spectrum 
obtained by
solving the Bogoliubov de Gennes equations 
as a function of $D$ 
for $A=12E_z$, $b=0.2 a_{z}$ and $\lambda=7$.
Solid and dashed lines in panel~(a) show the
real parts of the Bogoliubov
de Gennes excitation frequencies for $k=0$ and $1$,
while solid and dashed lines
in panel~(b) show those
for $k=2$ and $3$.
}\label{fig_bdglargelambda2}
\end{figure}

Figure~\ref{fig_bdglargelambda2} shows the Bogoliubov de Gennes excitation 
spectrum for $\lambda=7$, $A=12E_z$ and $b=0.2a_z$.
For this parameter combination, the energetically lowest lying
stationary GP solution deviates from the ``structureless Gaussian shape''
prior to collapse
and is instead characterized by a 
density whose maximum is located at $\rho>0$, i.e.,
the density is of type S$_{>0}$ prior to collapse
(see Fig.~\ref{fig_density} for a density profile of type S$_{>0}$
for a somewhat larger $\lambda$).
The Bogoliubov de Gennes excitation spectrum shown
in Figs.~\ref{fig_bdglargelambda2}(a) and \ref{fig_bdglargelambda2}(b)
is rich, with a series of crossings and avoided crossings.
Here, we focus on the large $D$ regime.
Figures~\ref{fig_bdglargelambda2}(a) and
\ref{fig_bdglargelambda2}(b) show that 
the lowest $k=3$ mode becomes
soft first, followed by the lowest $k=2$, $k=1$ and $k=0$ modes.
This indicates that
the 
collapse is triggered by a mode with non-vanishing azimuthal 
quantum number, similarly to the case with vanishing barrier 
height~\cite{rone07}.
The $D$ value, $D \approx 163$, at which the $k=3$ mode becomes soft
is about 16\% smaller than the $D$ value at which the stationary
GP equation first supports negative energy solutions. 
This implies an appreciable reduction of the S$_{>0}$ islands
shown in Figs.~\ref{fig_phasediagram} and \ref{fig_densityphase}.

It is worth emphasizing at this point that the fact that the
$k=3$ mode becomes soft first is not specific to the double-well
geometry considered here; in fact, decay triggered by the $k=3$
mode has also been found for 
pancake-shaped dipolar gases without barrier (see Ref.~\cite{rone07}
and below). While Fig.~\ref{fig_bdglargelambda2}
shows an example where the $k=3$ mode becomes soft first,
we find that the collapse
can---again, just as in the case of vanishing barrier~\cite{rone07}---also 
be triggered by other finite $k$ modes. For example,
for a somewhat smaller barrier height
but the same barrier width and aspect ratio as in 
Fig.~\ref{fig_bdglargelambda2}
(i.e., for $A=9E_z$, $b=0.2a_{z}$ and $\lambda=7$),
we find that the $k=2$ mode becomes soft first, followed by the
$k=3$ and $k=1$ modes (our calculations included modes $k=0$ through $4$).
In the following, we analyze the collapse triggered by the 
$k=3$ mode in more detail.

\begin{figure}
\vspace*{.2cm}
\includegraphics[angle=0,width=120mm]{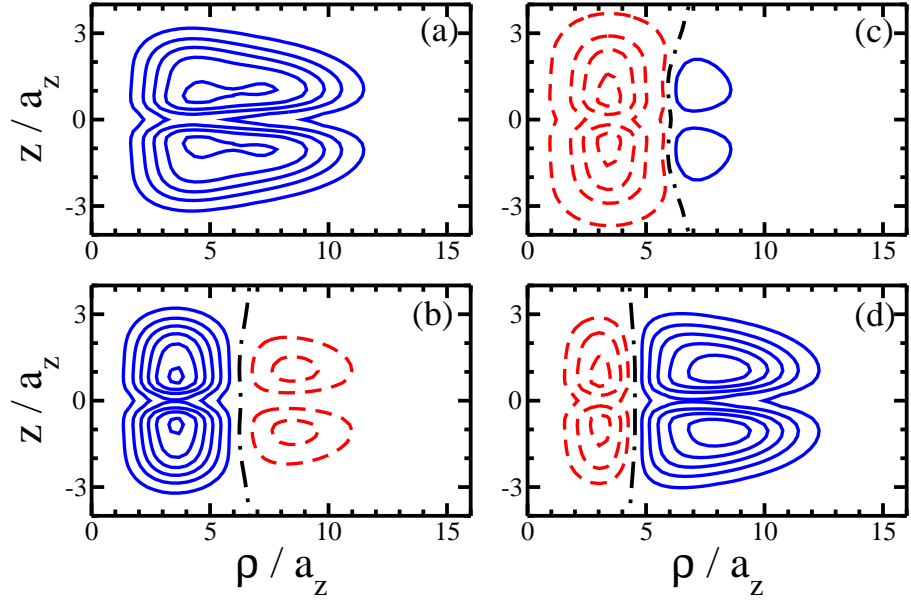}
\caption{(Color online)
Bogoliubov de Gennes eigenmodes $\bar{f}(\rho,z)$ corresponding to the 
two lowest
non-vanishing $k=3$ 
frequencies for $\lambda=7$,
$A=12E_z$ and $b=0.2 a_{z}$. 
Panels~(a) and (b)
show the eigenmodes corresponding to the
lowest and second lowest non-vanishing $k=3$ Bogoliubov de Gennes
frequencies for $D=119.1$ (i.e., prior to
the avoided crossing at $D\approx 140$), while
panels~(c) and (d)
show those for
$D=161.4$ (i.e., just prior to the lowest $k=3$ mode becoming soft).
The contours are chosen equidistant, with solid
and dashed lines corresponding
to positive and negative values
of $\bar{f}$. 
The dash-dotted lines indicate the nodal lines of 
$\bar{f}$.
}\label{fig_eigenmode4}
\end{figure}

Figure~\ref{fig_eigenmode4} shows the eigenmodes 
associated with the two lowest $k=3$ Bogoliubov
de Gennes frequencies for $D=119.1$ [Figs.~\ref{fig_eigenmode4}(a)
and \ref{fig_eigenmode4}(b)] and $D=161.4$
[Figs.~\ref{fig_eigenmode4}(c) and \ref{fig_eigenmode4}(d)].
Figure~\ref{fig_eigenmode4} 
shows that the eigenmode corresponding
to the lowest non-vanishing $k=3$ frequency has no nodal line prior to the
avoided crossing at $D\approx 140$ [see Fig.~\ref{fig_eigenmode4}(a)] 
but one nodal line, which can be
parametrized roughly by $\rho \approx 6a_z$ 
(or, equivalently by $\rho \approx 2.6 a_{\rho}$),
just prior to collapse [see Fig.~\ref{fig_eigenmode4}(c)].
Importantly, the eigenmode remains symmetric with respect to
$z=0$ even close to collapse and exhibits two equivalent extrema at
positive and negative $z$. This suggests that the collapse is associated 
with a total of six density peaks, three located on a ring of the 
red blood cell located in the left well and three
located on a ring of the red blood cell located in the right well.
The collapse can thus be characterized as local, 
with the local character arising from i) the angular roton like nature
of the instability
and ii) the presence of the Gaussian barrier.

\begin{figure}
\vspace*{.2cm}
\includegraphics[angle=0,width=120mm]{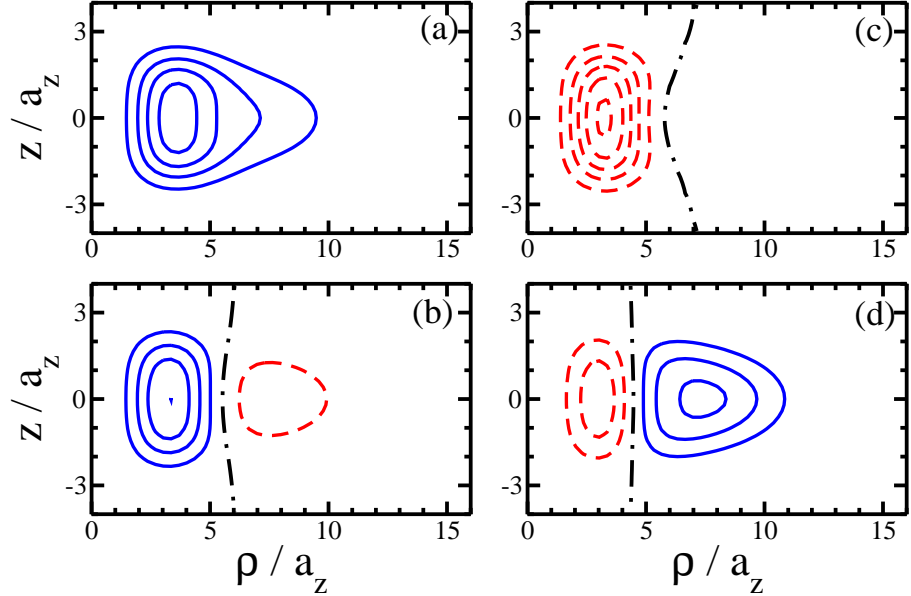}
\caption{(Color online)
Bogoliubov de Gennes eigenmodes $\bar{f}(\rho,z)$ corresponding to the 
two lowest
non-vanishing $k=3$ 
frequencies for $\lambda=7$,
$A=0$ and $b=0.2 a_{z}$. 
Panels~(a) and (b)
show the eigenmodes corresponding to the
lowest and second lowest non-vanishing $k=3$ Bogoliubov de Gennes
frequencies for $D=66.14$, while
panels~(c) and (d)
show those for
$D=81.49$ (i.e., just prior to the lowest $k=3$ mode becoming soft).
The contours are chosen equidistant, with solid
and dashed lines corresponding
to positive and negative values
of $\bar{f}$. 
The dash-dotted lines indicate the nodal lines of 
$\bar{f}$.
}\label{fig_eigenmode5}
\end{figure}

It is interesting to 
compare the eigenmodes for systems with vanishing and finite barrier
in the regime where the first Bogoliubov de Gennes mode becomes soft.
To this end,
Fig.~\ref{fig_eigenmode5} shows the eigenmodes 
corresponding to the two lowest non-vanishing $k=3$
frequencies
for two different $D$ values, i.e., for
$D = 66.14$ and $D=81.49$, and $\lambda=7$ and $A=0$.
The latter $D$ value corresponds to that investigated in
Fig.~2II of Ref.~\cite{rone07}. While we find that our excitation spectrum
for $A=0$ (not shown here) agrees with Fig.~2IIb of Ref.~\cite{rone07},
our eigenmode corresponding to the lowest non-vanishing 
$k=3$ frequency differs. In fact, we find that the eigenmode 
shown in Fig.~2IIc
for $D=81.49$~\cite{converttorho} corresponds to the second lowest and 
not to the 
lowest non-vanishing $k=3$ Bogoliubov de Gennes eigen frequency
as stated in Ref.~\cite{rone07}~\cite{agreementwithronen}.
The eigenmode shown in Fig.~\ref{fig_eigenmode5}(c) 
for $A=0$ and $D=81.49$ has a nodal line very similar to that
shown in Fig.~\ref{fig_eigenmode4}(c) for $A=12E_z$ and 
$D=161.4$, i.e., for a $D$ value
that is roughly twice as large as that for $A=0$.
The main difference between the two eigenmodes is that the latter is
characterized by extrema at $z \approx \pm a_{z}$ as opposed to $z=0$.

\section{Summary and Outlook}
\label{conclusion}
This paper presents a detailed mean-field analysis of a purely dipolar BEC 
in a cylindrically symmetric external confining potential with repulsive
Gaussian barrier centered at $z=0$.
The dipoles are assumed to be aligned along one of the symmetry axes of
the confining potential, which can be realized experimentally through
the application of an external field.
We have investigated the behaviors of the system as functions of the
dimensionless parameter $D$, which is defined as the product of the number of
particles and the square of the magnitude of the dipole moment $d$,
the aspect ratio $\lambda$ and, in a few selected cases, 
the height $A$ of the Gaussian barrier. 
Throughout, the
barrier width $b$ and the $s$-wave scattering length $a_s$ were kept
fixed at $b=0.2a_z$ and $a_s=0$, respectively.
The energetics and the density profiles obtained 
by solving the stationary GP equation have been discussed
and the onset of the mechanical instability
has been analyzed. Additional insights were gained from
a dynamical stability analysis, which is based on
time-evolving a given initial state or
on the Bogoliubov de Gennes excitation spectrum.
The latter was complemented by a detailed analysis of selected Bogoliubov
de Gennes eigenmodes.

For sufficiently small aspect ratios, i.e., for cigar-shaped 
harmonic traps in which the dipoles are aligned along the weak confinement
direction, the energetically lowest lying stationary ground
state solution is either symmetric or symmetry-broken. As in the
case of $s$-wave interacting BECs, the appearance of these solutions
can be explained qualitatively within a two-mode model.
For a critical mean-field strength $D_{\mathrm{cr}}$, 
the system becomes unstable.
The Bogoliubov de Gennes framework reveals that the collapse occurs globally,
i.e., within a single well, with the decay being triggered 
by a mode with vanishing
azimuthal quantum number.

As the aspect ratio $\lambda$ increases, the region of the
phase diagram where the energetically lowest lying stationary GP
solution is symmetry-broken vanishes. For sufficiently large aspect ratio
$\lambda$, the symmetric solutions fall into one of two classes:
The system's density is characterized by a density maximum at $\rho=0$
(this is referred to as S$_0$ type)
or by a density maximum at $\rho>0$
(this is referred to as S$_{>0}$ type). 
The latter class of solutions occupies a fairly small region of the phase diagram
and only occurs near the dynamical instability line and only for
certain aspect ratios $\lambda$ and barrier heights $A$.
For a barrier height of $A=12E_z$ and $\lambda=7$, e.g., 
the solution is of type S$_{>0}$ prior to collapse.
Correspondingly, we find that the
collapse occurs locally through a mode with finite
azimuthal quantum number ($k=3$), 
with density spikes emerging in six different regions of the
trap. Three of these density peaks grow 
in the left well and three in the right well. Furthermore, 
each of the three density peaks lies on a ring that 
is associated with the 
density maximum of the ground state solution 
at $\rho>0$. The instability can, as in the case of a vanishing
Gaussian barrier, be characterized as 
an angular roton instability. We note, however, that the
radial degrees of freedom also play a role, i.e., that the Bogoliubov 
de Gennes eigenmodes prior to collapse contain
radial nodal lines. 
For pancake-shaped
trapping geometries (i.e., for $\lambda>1$),
this paper primarily explored the
$(D,\lambda)$ parameter space for fixed $A$.
We find that the double well system exhibits a number
of novel and rich
stability characteristics as the barrier height $A$ is varied;
these studies will be reported on in a forthcoming article~\cite{asad09}.

Our theoretical predictions for purely dipolar BECs presented in this paper
can be tested experimentally by loading a 
dipolar BEC such as a Cr BEC into a double well potential
and by tuning the $s$-wave scattering length to zero 
through the application
of an external magnetic field in the vicinity of a 
magnetic Feshbach resonance. 
Following the spirit of the double-well experiments 
for $s$-wave interacting BECs (see, e.g., Ref.~\cite{albi05}),
it should be possible to study the transition
from the Josephson tunneling regime to the macroscopic 
quantum self-trapping regime by
loading the double well system with varying number of particles
and varying population difference in the left
and right wells. 
For pancake-shaped harmonic confinement,
we suggest 
an
experimental sequence
that would allow the stability
lines discussed in the context
of Fig.~\ref{fig_densityphase}
to be probed.
We 
suggest to increase the radial trapping frequency $\omega_{\rho}$ 
(and to thus decrease $\lambda$)
for fixed $A$
and to monitor the loss of atoms from the trap.
This scenario corresponds to approaching 
the instability line in Fig.~\ref{fig_densityphase} vertically from above.
By repeating 
this
experiment
for condensates with varying number of particles,
the different collapse mechanisms associated with the S$_{0}$ regions and the 
S$_{>0}$ islands could be probed (see also Ref.~\cite{wilsonmostrecent}).

In the future,
it will
be interesting to investigate
how the behaviors of the system change as a 
function of the ``spacing'' between the left and the
right well, i.e., as a function of the barrier widths $b$. 
The present study covers the regime where the 
``spacing'' between the 
left and the right well is comparatively small, i.e., where it is
comparable to the
axial confinement length, and where
the system is described by one macroscopic wave function.
Our approach for pancake-shaped confinement with Gaussian barrier
is distinctly different from
other recent studies of multi-layer (quasi-)two-dimensional dipolar 
BECs~\cite{wang08,kobe09},
which assume that the dipoles in neighboring wells
feel each other but that the distance
between the neighboring wells is so large that each dipole can be
assigned to a specific well.
In this case, the system has been discretized, leading to a coupled
set of equations that have been solved self-consistently.
It will be interesting to extend the present study of pancake-shaped
two-well systems to a regime where comparisons with a 
discretized description become meaningful.
It will also be interesting to extend the present work to multi-well 
traps. In the regime of small aspect ratio, e.g.,
a three- or four-well system
might lead to interesting dynamics that can be controlled by varying the 
onsite and the offsite interactions. 
In this case, a multi-mode analysis suggests
itself as a first starting point.

Support by the NSF through
grants PHY-0555316
and PHY-0855332
is gratefully acknowledged.


\begin{thebibliography}{10}
\bibitem{bara08}
M.~A. Baranov, Physics Reports {\bf 464},  71  (2008).

\bibitem{laha08a}
T. Lahaye, C. Menotti, L. Santos, M. Lewenstein, 
and T. Pfau, arXiv:0905.0386 (2009).

\bibitem{grie05}
A. Griesmaier, J. Werner, S. Hensler, J. Stuhler, and
T. Pfau, Phys. Rev. Lett. {\bf 94},  160401  (2005).

\bibitem{stuh05}
J. Stuhler, A. Griesmaier, T. Koch, M. Fattori,
T. Pfau, S. Giovanazzi, P. Pedri, and L. Santos, 
Phys. Rev. Lett. {\bf 95},  150406  (2005).

\bibitem{wern05}
J. Werner, A. Griesmaier, S. Hensler, J. Stuhler,
T. Pfau, A. Simoni and E. Tiesinga, Phys. Rev. Lett. {\bf 94},  183201  (2005).

\bibitem{koch08}
T. Koch,  T. Lahaye,
J. Metz, B.  Fr\"ohlich, 
A. Griesmaier, and T. Pfau, Nature Physics {\bf 4},  218  (2008).

\bibitem{laha08}
T. Lahaye, J. Metz, B. Fr\"ohlich,
T. Koch, M. Meister, A. Griesmaier, T. Pfau, H. Saito, 
Y. Kawaguchi, and M. Ueda, Phys. Rev. Lett. {\bf 101},  080401  (2008).

\bibitem{sant00}
L. Santos, G.~V. Shlyapnikov, P. Zoller, and M. Lewenstein, Phys. Rev. Lett.
  {\bf 85},  1791  (2000).

\bibitem{yi00}
S. Yi and L. You, Phys. Rev. A {\bf 61},  041604(R)  (2000).

\bibitem{yi01}
S. Yi and L. You, Phys. Rev. A {\bf 63},  053607  (2001).

\bibitem{gora02}
K. Goral and L. Santos, Phys. Rev. A {\bf 66},  023613  (2002).

\bibitem{eber05}
C. Eberlein, S. Giovanazzi, and D.~H.~J. O'Dell, Phys. Rev. A {\bf 71},  033618
   (2005).

\bibitem{bort06}
D.~C.~E. Bortolotti, S. Ronen, J.~L. Bohn, and D. Blume, Phys. Rev. Lett. {\bf
  97},  160402  (2006).

\bibitem{fisc06}
U.~R. Fischer, Phys. Rev. A {\bf 73},  031602(R)  (2006).

\bibitem{rone07}
S. Ronen, D.~C.~E. Bortolotti, and J.~L. Bohn, Phys. Rev. Lett. {\bf 98},
  030406  (2007).

\bibitem{dutt07}
O. Dutta and P. Meystre, Phys. Rev. A {\bf 75},  053604  (2007).

\bibitem{metz09}
J. Metz, T. Lahaye, B. Fr\"ohlich, A. Griesmaier,
T. Pfau, H. Saito, Y. Kawaguchi, and M. Ueda, 
New J. Phys. {\bf 11},  055032  (2009).

\bibitem{giov02}
S. Giovanazzi, D. O'Dell, and G. Kurizki, Phys. Rev. Lett. {\bf 88},  130402
  (2002).

\bibitem{sant03}
L. Santos, G.~V. Shlyapnikov, and M. Lewenstein, Phys. Rev. Lett. {\bf 90},
  250403  (2003).

\bibitem{coop05}
N.~R. Cooper, E.~H. Rezayi, and S.~H. Simon, Phys. Rev. Lett. {\bf 95},  200402
   (2005).

\bibitem{zhan05}
J. Zhang and H. Zhai, Phys. Rev. Lett. {\bf{95}}, 200403 (2004).

\bibitem{yi06}
S. Yi and H. Pu, Phys. Rev. A {\bf 73},  061602(R)  (2006).

\bibitem{odel07}
D.~H.~J. O'Dell and C. Eberlein, Phys. Rev. A {\bf 75},  013604  (2007).

\bibitem{wils09}
R.~M. Wilson, S. Ronen, and J.~L. Bohn, Phys. Rev. A {\bf 79},  013621  (2009).

\bibitem{gora02a}
K. G\'oral, L. Santos, and M. Lewenstein, Phys. Rev. Lett. {\bf 88},  170406
  (2002).

\bibitem{dams03}
B. Damski, L. Santos, E. Tiemann,
M. Lewenstein, S. Kotochigova, P. Julienne, and P. Zoller, 
Phys. Rev. Lett. {\bf 90},  110401  (2003).

\bibitem{meno07}
C. Menotti, C. Trefzger, and M. Lewenstein, Phys. Rev. Lett. {\bf 98},  235301
  (2007).

\bibitem{sadl06}
L.~E. Sadler, J. M. Higbie, S. R. Leslie, M. Vengalatorre,
and D. M. Stamper-Kurn, Nature {\bf 443},  312  (2006).

\bibitem{veng08}
M. Vengalatorre, S.~R. Leslie, J. Guzman, and D.~M. Stamper-Kurn, Phys. Rev.
  Lett. {\bf 100},  170403  (2008).

\bibitem{fatt08}
M. Fattori, G. Roati, B. Deissler, C. D'Errico, M. Zaccanti,
M. Jona-Lasinio, L. Santos, M. Inguscio, and G. Modugno, 
Phys. Rev. Lett. {\bf 101},  190405  (2008).

\bibitem{poll09}
S.~E. Pollack, D. Dries, M. Junker, Y. P. Chen,
T. A. Corcovilos, and R. G. Hulet, 
Phys. Rev. Lett. {\bf 102},  090402  (2009).

\bibitem{ni08}
K.-K. Ni, S. Ospelkaus, M. H. G. {de Miranda},
A. Pe'er, B. Neyenhuis, J. J. Zirbel, 
S. Kotochigova, P. S. Julienne, D. S. Jin, and J. Ye, 
Science {\bf 322},  231  (2008).

\bibitem{deig08}
J. Deiglmayr, A. Grochola, M. Repp, K. Mortlbauer,
C. Gluck, J. Lange, O. Dulieu, R. Wester, and M. Weidem\"uller, 
Phys. Rev. Lett. {\bf 101},  133004  (2008).

\bibitem{danz08}
J.~G. Danzl, E. Haller, M. Gustavsson, M. J. Mark,
R. Hart, N. Bouloufa, O. Dulieu, H. Ritsch, and H.-C. N\"agerl, 
Science {\bf 321},  1062  (2008).

\bibitem{mari98}
M. Marinescu and L. You, Phys. Rev. Lett. {\bf 81},  4596  (1998).

\bibitem{lewe07}
M. Lewenstein, A.  Sanpera, V. Ahufinger, B. Damski, 
A. {Sen De}, and U. Den, 
Adv. Phys. {\bf 56},  243  (2007).

\bibitem{bloc08}
I. Bloch, J. Dalibard, and W. Zwerger, Rev. Mod. Phys. {\bf 80},  885  (2008).

\bibitem{aubr96}
S. Aubry, S. Flach, K. Kladko, and E. Olbrich, Phys. Rev. Lett. {\bf 76},  1607
   (1996).

\bibitem{smer97}
A. Smerzi, S. Fantoni, S. Giovanazzi, and S.~R. Shenoy, Phys. Rev. Lett. {\bf
  79},  4950  (1997).

\bibitem{ande98a}
B.~P. Anderson and M.~A. Kasevich, Science {\bf 282},  1686  (1998).

\bibitem{ragh99}
S. Raghavan, A. Smerzi, S. Fantoni, and S.~R. Shenoy, Phys. Rev. A {\bf 59},
  620  (1999).

\bibitem{giov00a}
S. Giovanazzi, A. Smerzi, and S. Fantoni, Phys. Rev. Lett. {\bf 84},  4521
  (2000).

\bibitem{cata01}
F.~S. Cataliotti, S. Burger, C. Fort, P. Maddaloni,
F. Minardi, A. Trombettoni, A. Smerzi, and M. Inguscio, 
Science {\bf 293},  843  (2001).

\bibitem{albi05}
M. Albiez, R. Gati, J. Folling, S. Hunsmann,
M. Cristiani, and M. K. Oberthaler, Phys. Rev. Lett. {\bf 95},  010402  (2005).

\bibitem{anan06}
D. Ananikian and T. Bergeman, Phys. Rev. A {\bf 73},  013604  (2006).

\bibitem{gati07}
R. Gati and M.~K. Oberthaler, J. Phys. B {\bf 40},  61{R}  (2007).

\bibitem{este08}
J. Esteve, C. Gross, A. Weller, S. Giovanazzi,
and M. K. Oberthaler, Nature {\bf 455},  1216  (2008).

\bibitem{shin04}
Y. Shin, M. Saba,
T. A. Pasquini, W. Ketterle, D. E. Pritchard,
and A. E. Leanhardt, Phys. Rev. Lett. {\bf 92},  050405  (2004).

\bibitem{xion09}
B. Xiong, J. B. Gong, H. Pu, W. Z. Bao, and B. W. Li, 
Phys. Rev. A {\bf 79},  013626  (2009).

\bibitem{wang08}
D.-W. Wang and E. Demler, arXiv:0812.1838  .

\bibitem{klaw09}
M. Klawunn and L. Santos, Phys. Rev. A {\bf 80},  013611  (2009).

\bibitem{kobe09}
P. K\"oberle and G. Wunner, arXiv:0908.1009  (2009).

\bibitem{gora00}
K. Goral, K. Rzazewski, and T. Pfau, Phys. Rev. A {\bf 61},  051601(R)  (2000).

\bibitem{rone06a}
S. Ronen, D.~C.~E. Bortolotti, and J.~L. Bohn, Phys. Rev. A {\bf 74},  013623
  (2006).

\bibitem{modu03}
M. Modugno, L. Pricoupenko, and Y. Castin, Eur. Phys. J. D {\bf 22},  235
  (2003).

\bibitem{dalf98}
F. Dalfovo, S. Giorgini, L.~P. Pitaevskii, and S. Stringari, Rev. Mod. Phys.
  {\bf 71},  463  (1999).

\bibitem{dalf97}
F. Dalfovo, S. Giorgini, M. Guilleumas,
L. P. Pitaevskii, and S. Stringari, Rev. Phys. A {\bf 56},  3840  (1997).

\bibitem{webb74}
R.~A. Webb, R.~L. Kleinberg, and J.~C. Wheatley, Phys. Rev. Lett. {\bf 33},
  145  (1974).

\bibitem{levy07}
S. Levy, E. Lahoud, I. Shomroni, and J. Steinhauer, Nature {\bf 449},  579
  (2007).

\bibitem{wils08}
R.~M. Wilson, S. Ronen, J.~L. Bohn, and H. Pu, Phys. Rev. Lett. {\bf 100},
  245302  (2008).


\bibitem{converttorho}
$d^2(N-1)=30.8E_{\rho}a_{\rho}^3= 30.8 \sqrt{\lambda} E_{z}a_{z}^3$, which
  implies that $D=81.49$ corresponds to $d^2(N-1)=30.8 E_{\rho}a_{\rho}^3$ for
  $\lambda=7$.

\bibitem{agreementwithronen}
This has been confirmed by S.~Ronen in a private communication 
(August 2009).

\bibitem{asad09}
M. Asad-uz-Zaman and D. Blume, in preparation.

\bibitem{wilsonmostrecent}
R.~M. Wilson, S. Ronen and J.~L. Bohn, Phys. Rev. A {\bf{80}}, 023614 (2009).

\end{thebibliography}
\end{document}